\documentclass{aastex6}

\usepackage{multirow}

\fullcollaborationName{The Friends of AASTeX Collaboration}

\begin{document}

\title{The Gould's Belt Distances Survey (GOBELINS) III. \\ The distance to the Serpens/Aquila Molecular Complex}

\author{ Gisela N. Ortiz-Le\'on\altaffilmark{1}, 
Sergio A.\ Dzib\altaffilmark{2},
Marina A.\ Kounkel\altaffilmark{3},
Laurent Loinard\altaffilmark{1,2}, 
Amy J.\ Mioduszewski\altaffilmark{4},
Luis F.\ Rodr\'{\i}guez\altaffilmark{1},
Rosa M.\ Torres\altaffilmark{5},
Gerardo Pech\altaffilmark{1},
Juana L.\ Rivera\altaffilmark{1},
Lee Hartmann\altaffilmark{3},
Andrew F.\ Boden\altaffilmark{6},
Neal J.\ Evans II\altaffilmark{7},
Cesar Brice\~no\altaffilmark{8},
John J.\ Tobin\altaffilmark{9,10}, and
Phillip A.~B.\ Galli\altaffilmark{11,12}
}

\email{g.ortiz@crya.unam.mx}

\altaffiltext{1}{Instituto de Radioastronom\'ia y Astrof\'isica, 
Universidad Nacional Aut\'onoma de M\'exico,
Morelia 58089, M\'exico}
\altaffiltext{2}{Max Planck Institut f\"ur Radioastronomie, Auf dem H\"ugel 69, 
D-53121 Bonn, Germany}
\altaffiltext{3}{Department of Astronomy, University of Michigan, 500 Church Street, 
Ann Arbor, MI 48105,  USA}
\altaffiltext{4}{National Radio Astronomy Observatory, Domenici Science Operations 
Center, 1003 Lopezville Road, Socorro, NM 87801, USA}
\altaffiltext{5} {Centro Universitario de Tonal\'a, Universidad de Guadalajara,
Avenida Nuevo Perif\'erico No.\ 555, Ejido San Jos\'e Tatepozco, C.P.
48525, Tonal\'a, Jalisco, M\'exico. }
\altaffiltext{6}{Division of Physics, Math and Astronomy, California Institute of Technology, 
1200 East California Boulevard, Pasadena, CA 91125, USA} 
\altaffiltext{7}{Department of Astronomy, The University of Texas at Austin, 
2515 Speedway, Stop C1400, Austin, TX 78712-1205, USA}
\altaffiltext{8}{Cerro Tololo Interamerican Observatory, Casilla 603, La Serena, Chile}
\altaffiltext{9}{Homer L. Dodge Department of Physics and Astronomy, University of Oklahoma, 440 W.
Brooks Street, Norman, OK 73019, USA}
\altaffiltext{10}{Leiden Observatory, PO Box 9513, NL-2300 RA, Leiden, The Netherlands}
\altaffiltext{11}{Instituto de Astronomia, Geof\'isica e Ci\^encias Atmosf\'ericas, Universidade de S\~ao Paulo,
Rua do Mat\~ao 1226, Cidade Universit\'aria, S\~ao Paulo, Brazil }
\altaffiltext{12}{Univ.\ Grenoble Alpes, IPAG, 38000, Grenoble, France}

\begin{abstract}
We report on new distances and proper motions to seven stars across the Serpens/Aquila complex. The observations were obtained as part of the Gould's Belt Distances Survey (GOBELINS) project between September 2013 and  April 2016 with the Very Long Baseline Array (VLBA). One of our targets is the proto-Herbig AeBe object EC~95, which is a binary system embedded in the Serpens Core.  For this system, we combined the GOBELINS observations with previous VLBA data to cover a total period of ~8 years, and derive the orbital elements and an updated source distance.  The individual distances to sources in the complex  are fully consistent with each other,  and the mean value corresponds to a distance  of $436.0\pm9.2$~pc for the Serpens/W40 complex. Given this new evidence, we argue that Serpens Main, W40 and Serpens South are physically associated and form a single cloud structure. 

\end{abstract}


\section{Introduction}\label{sec:intro}


The Serpens molecular cloud is a region rich in low-mass star formation selected for observations as part of the Gould's Belt Distances Survey (GOBELINS; Ortiz-Le\'on et al.\ 2016). There are two smaller regions, of $\sim 1$~deg$^2$ in size, associated with this cloud: Serpens Main and Serpens South. Serpens Main  \citep[centered on R.A.\ $18^{\rm h}29^{\rm m}00^{\rm s}$, Dec.\ $+00^{\rm o}30'00''$;][]{Eiroa_2008} consists of three prominent  sub-regions; namely the Serpens core,  Serpens G3-G6 and VV Serpentis.  Its northernmost sub-region is the Serpens core \citep[also called Serpens North or Cluster A;][]{Harvey_2006}, a cluster of YSOs deeply embedded with extinction exceeding 40 mag in the visual. This sub-region  has  numerous observations from X-rays to the submillimeter that have revealed a large population of protostars \citep[e.g.,][]{Kaas_2004, Eiroa_2005, Harvey_2006, Harvey_2007, Winston_2007, Winston_2009, Oliveira_2010}. Serpens G3-G6 \citet{Cohen_1979}, also  referred to as Cluster B,  was identified by \cite{Harvey_2006} as a cluster of star formation harboring many previously unknown young stellar objects (YSOs). Finally, VV Serpentis is the southernmost sub-region associated to the eponymous star. Presently, the most extensive study of the young stellar population in Serpens Main was conducted by the Spitzer Legacy Program ``From Molecular Cores to Planet-Forming Disks'' \citep[c2d;][]{Evans_2003}, where more than two hundred Class 0 to Class III YSOs associated with IR excess were identified in an area of  0.85 deg$^2$ \citep{Dunham_2015}. Serpens South (centered on R.A.\ $18^{\rm h}30^{\rm m}00^{\rm s}$, Dec.\ $-02^{\rm o}02'00''$, i.e.\ at an angular distance of $\sim 3^{\rm o}$ to the south of Serpens Main) was discovered by \cite{Gutermuth_2008}. Since then, it has received a lot of attention because of the large number of extremely young objects that it contains. 
It shows an unusually large fraction of protostars \citep{Gutermuth_2008}, presenting an excellent laboratory to study the earliest stages of star formation. 

\begin{figure*}[!ht]
\begin{center}
 \includegraphics[width=0.9\textwidth,angle=0]{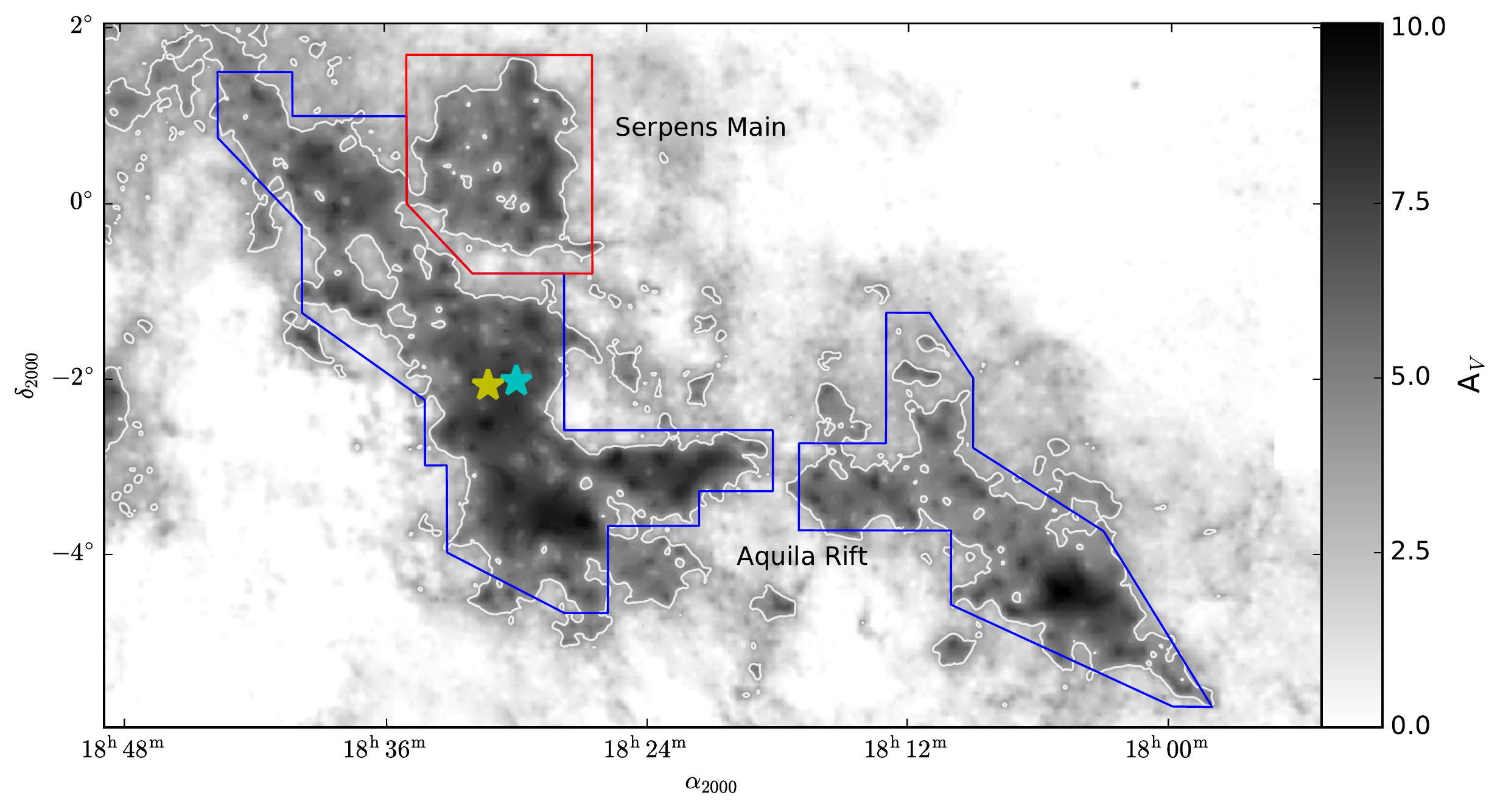}
\end{center}
 \caption{Extinction map of the Serpens and Aquila star-forming regions obtained as part of the COMPLETE project, based on the STScI Digitized Sky Survey data \citep{Cambresy_1999}. Red and blue polygons mark the structures corresponding to Serpens Main and the Aquila Rift, respectively, while cyan and yellow stars indicate the center of the W40 and Serpens South regions. The white contour indicates an $A_V$ of 4. }
\label{fig:ser_aq}
\end{figure*}

To the east of Serpens South, at R.A.\ $\sim 18^{\rm h}31^{\rm m}29^{\rm s}$, Dec.\ $-02^{\rm o}05'24''$, lies the W40 complex, named after the  H II region, also known as Sharpless 2-64 \citep{Smith_1985, Vallee_1987}. This complex shows evidence for ongoing star formation, since it contains dense molecular cores \citep{Dobashi_2005}, millimeter-wave sources \citep{Molinari_1996, Maury_2011}, and YSOs \citep{Kuhn_2010, Rodriguez_2010,  Mallick_2013}. There is also a cluster of massive stars which ionizes the H II region 
\citep{Smith_1985, Shuping_2012}. Both Serpens South and W40 belong to a larger complex of molecular clouds collectively known as the {\it Aquila Rift}, a large elongated feature seen in 2MASS extinction maps \citep{Bontemps_2010}. The Aquila Rift was one of the  clouds targeted by the {\it Herschel} \citep{Andre_2010,Konyves_2015} and {\it Spitzer} \citep{Dunham_2015} Gould Belt Surveys, which revealed hundreds of YSOs all across the complex.  Figure \ref{fig:ser_aq} shows the  location  of the Serpens Main region, as well as the position of W40 and Serpens South  within the Aquila Complex. We note that, although Serpens and the Aquila Rift do not formally belong to the Gould's Belt, 
they are usually included in Gould Belt Surveys because of their star-formation activity,  and because they were previously thought to be closer to the Sun. 

\subsection{The distance to the clouds in Serpens/Aquila}\label{sec:intro_dist}

The distances to the different regions in the Serpens/Aquila Complex have been a matter of controversy. For Serpens Main, there is an ample range of distances reported in the literature,  from $245\pm30$~pc \citep{ChavarriaK_1988} to $650\pm180$~pc \citep{Zhang_1988}. Most of these estimates are indirect, since  they are often based on spectroscopic parallaxes and extinction measurements. \cite{Winston_2010} constructed the X-ray luminosity function of  the Serpens cluster using different distances to calculate the X-ray luminosity, and fitted the data with the distribution determined by \cite{Feigelson_2005} for Orion, IC 348 and NGC 1333. The best fit to the data was found to be at a distance to Serpens of $360^{+22}_{-13}$~pc. The only direct measurement of the distance to Serpens Main has been obtained by \cite{Dzib_2010} and \cite{Dzib_2011} from Very Long Baseline Interferometry (VLBI) trigonometric parallax of the YSO EC 95 associated with the Serpens Core. These authors derived a distance to the Serpens Core of $415\pm5$~pc and a mean distance to the Serpens cloud of $415\pm25$~pc. Later,  they updated the distance to the  Core to $429\pm2$~pc. 
However, the usually adopted distance  for Serpens Main and the Aquila Rift as well is $259\pm37$~pc, which was derived by \cite{Straizys_1996} from photometry of $\sim 100$ optically visible stars, $18$ of which belong to Serpens Main.  In a more recent paper, \cite{Straizys_2003} used 80 stars from their original sample,  as well as 400 other stars, to measure the distance to the front edge of the dark clouds (the extinction wall) in the Serpens/Aquila complex. They placed this wall at  $225\pm55$~pc, and suggested that the cloud is about  80 pc deep.  

As we mentioned earlier, W40 and Serpens South are embedded within the Aquila Rift.
Estimates of the distance to W40 seem to favor values between  455 and 600 pc \citep{Kolesnik_1983, Shuping_2012}, which suggests this cloud lies somewhat further away than the extinction wall of the  Aquila Rift. So far, there are no distance measurements to sources in Serpens South, but many authors argue that the region is at the same distance as Serpens Main, and adopt either 260 or 429 pc \citep[e.g.,][]{Gutermuth_2008, Maury_2011, Plunkett_2015, Kern_2016, Heiderman_2015}.  It has also been argued that W40 and Serpens South belong to the same continuous extinction feature and should be part of the same complex, likely at the same distance \citep{Maury_2011}.  

\bigskip

In this paper, we report on the distance to 3 stars in the Serpens cloud core and 4 objects in the W40 cluster.
The observations were obtained as part of the GOBELINS project (Ortiz-Le\'on et al.\ 2016) with the Very Long Baseline Array (VLBA). We describe our targets and observations in  Section \ref{sec:targets}. 
The astrometry of our sources is given in Section \ref{sec:results}.
Finally, we discuss our findings in Section \ref{sec:discussion} and provide a summary in Section \ref{sec:summary}.

\section{Target selection and Observations}\label{sec:targets}

While both thermal and non-thermal processes produce radio emission in young stars, only brightness temperatures $\gtrsim10^6$~K will be detectable on VLBI baselines  \citep{Thompson_2007}, which
limits VLBI observations to non-thermal radiation.
Thus, our targets consist of young stars with potentially non-thermal radio emission. This kind of emission is expected to be produced in the coronae of magnetically active stars by energetic electrons gyrating around the magnetic field lines \citep{Feigelson_1999}. 

\begin{figure*}[!ht]
\begin{center}
 \includegraphics[width=0.9\textwidth,angle=0]{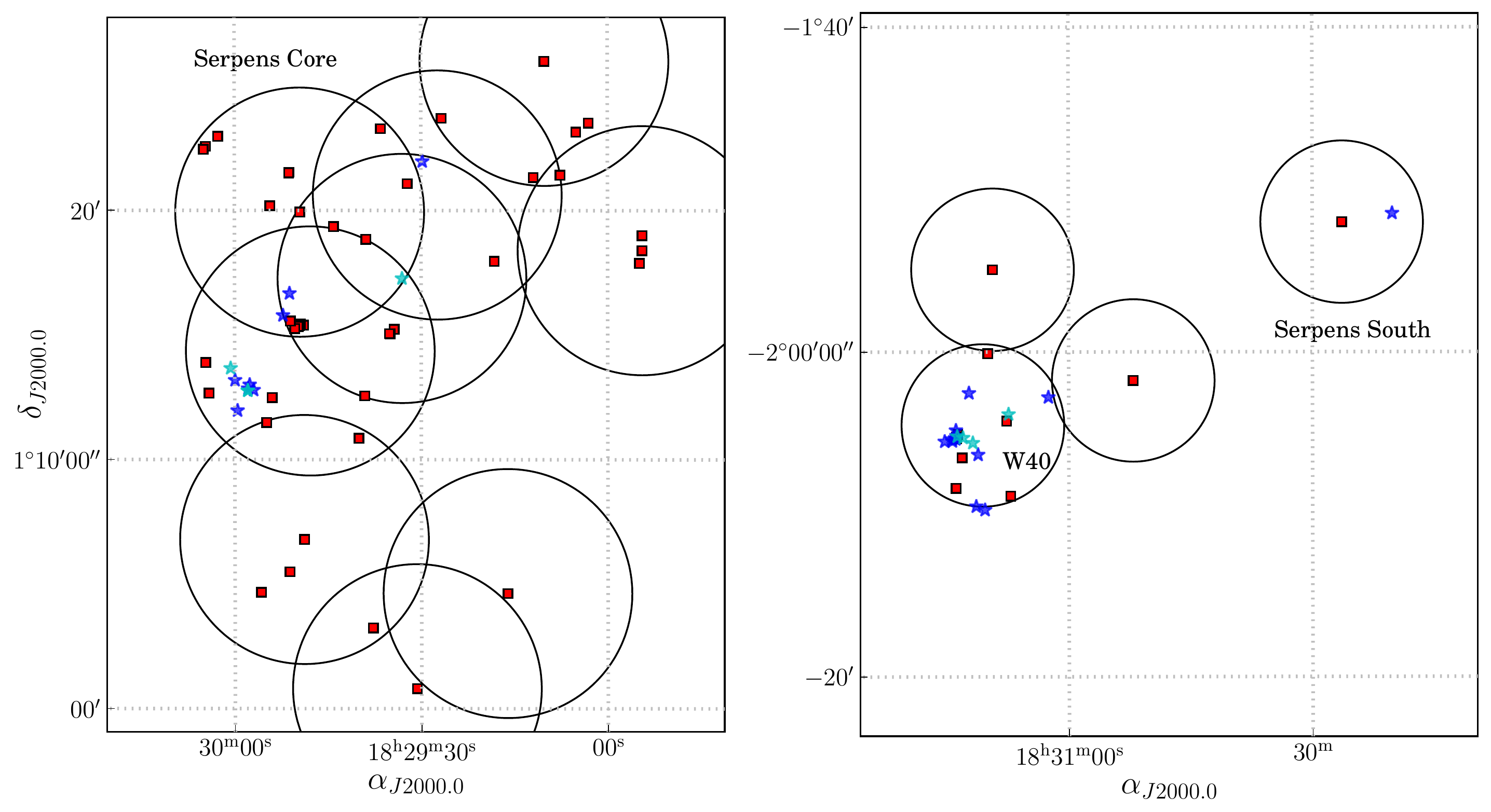}
\end{center}
 \caption{Spatial distribution of observed sources in the Serpens Core (left) and W40/Serpens South (Right). Blue and cyan stars correspond  to known YSOs and YSOs with a distance estimation provided in this paper, respectively. Red squares mark the positions of other unclassified observed sources. }
\label{fig:pointings}
\end{figure*}

In \citet{Ortiz_15}, we reported on deep radio observations carried out with 
the Karl Jansky Very Large Array (VLA) of  three of the most prominent regions in the Serpens/Aquila Complex, namely, the Serpens Core, W40 and Serpens South. 
A total of 18 possible targets (known or candidate YSOs) for VLBA astrometry were identified across these three regions, based on their compactness, negative spectral index and/or variability.
The VLBA was pointed at the positions of the 18 candidates, however we also correlated (i.e., changed the phase center of the correlation) at the positions of other sources which lay in the primary beam of the individual VLBA telescopes (of $10'$ in size at 5~GHz).  This provided an additional 63 sources, of which 3 turned out to be YSOs with detectable non-thermal radio emission.  

We refer the reader to Ortiz-Le\'on et al.\ (2016) for a detailed description of our observing approach. Briefly, the VLBA observations  of GOBELINS were taken between 2013 September and 2016 April at $\nu=$~4.9 or 8.3~GHz (C- and X-band, respectively). The data were recorded in dual polarization mode with 256 MHz of bandwidth in each polarization, covered by 8 separate 32-MHz intermediate frequency (IF) channels. VLBA  project codes, observing dates, pointing positions, and corresponding observing bands are given in Table \ref{tab:obs}. Several sets of phase calibrators were chosen according to their angular separations relative to target positions and used for multi-source phase referencing. The corresponding sets of calibrators for each pointing position (target) are listed in Table \ref{tab:point}. One or two targets were observed in each observing session. These consisted of cycles alternating between the target(s) and the main phase calibrator: $\{$target --- calibrator$\}$ for single-target sessions, and $\{$target 1 --- calibrator --- target 2 --- calibrator$\}$ for those sessions where two targets were observed simultaneously. The secondary calibrators were observed every $\sim50$~minutes. The total integration time for each target was
$\sim1.6$~hours in projects that observed at 8.3 GHz, and $\sim1$~hour at 4.9 GHz.  Geodetic-like blocks, consisting of observations of many calibrators over a wide range of elevations, were taken  before and after each session. These were observed with 512 MHz total bandwidth covered by 16 IFs and centered at $\nu=$~4.6 and 8.1~GHz for projects observing at C- and X-band, respectively.  

Data reduction was performed using AIPS \citep{Greisen_2003}, following the strategy described in Ortiz-Le\'on et al.\ (2016).  Calibrated visibilities were imaged using a pixel size of 50--100~$\mu$as and pure natural weighting. Typical angular resolutions were 4 mas  $\times$ 2 mas ($\sim 1.3$ AU at a distance of 429~pc) at 4.9 GHz and 3 mas $\times$ 0.9 mas ($\sim 0.8$ AU) at  8.3 GHz. Noise levels were typically 30 and $38~\mu{\rm Jy~beam}^{-1}$ at C- and X-band, respectively. 

In addition, we will use of data from VLBA projects BL155 and BL160 (P.I.: L.\ Loinard) and BD155 (P.I.: S.\ Dzib) which were designed to only observe the source EC~95 between 2007, December and 2016, January at $\nu=8.4$~GHz. The images corresponding to these old observations have noise levels of $76~\mu{\rm Jy~beam}^{-1}$. 

\section{Results}\label{sec:results}


As mentioned earlier, we observed a total of 81 sources in the
Serpens/Aquila region. Their spatial distribution  is shown in Figure \ref{fig:pointings}, while source VLA coordinates, names, source types, fluxes and brightness temperatures, $T_b$, are given in the first eight columns in Table \ref{tab:sources}. 
Out of the total observed sources, 30 have been firmly detected. These are sources detected in several epochs, with at least one detection  at $5\sigma$, or sources detected just in one epoch but at $6\sigma$, where $\sigma$ is the {\it rms} noise measured in the images. 
All sources show $T_b > 10^6$~K, consistent with the brightness temperature expected for non-thermal emission.  

%

\begin{figure*}[!ht]
\begin{center}
 {\includegraphics[width=0.5\textwidth,angle=0]{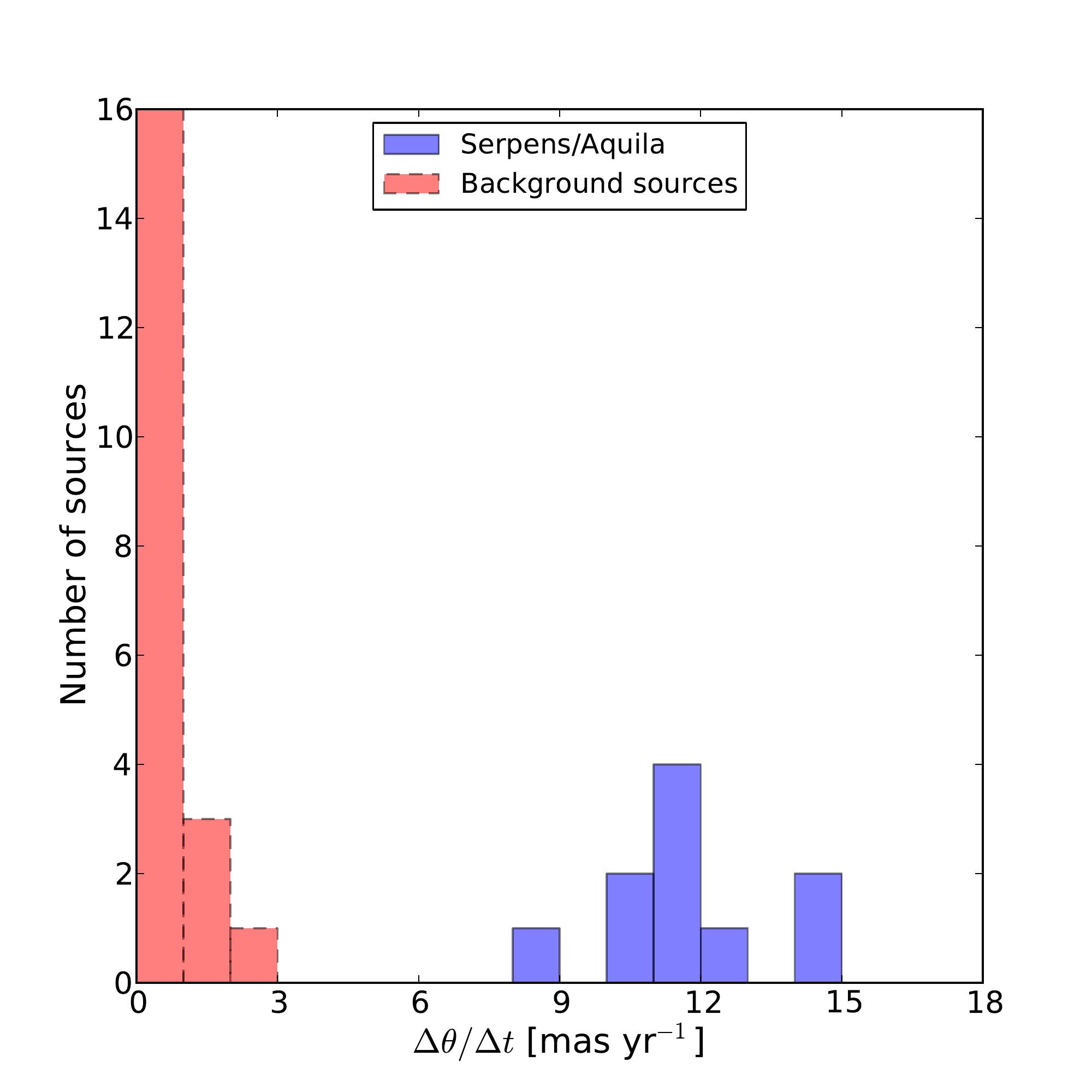}}
\end{center}
 \caption{
 Histogram of position change rate for all sources detected at least twice toward Serpens/Aquila. The sources previously identified as members of the complex are shown as a blue histogram. These are 10 sources: 5 single YSOs, the 4 components of the two binary systems and the B1V star. The source KGF~138 is not shown because it  has been detected only once. Other sources, whose classification is unknown in the literature, are shown as a red histogram.   
 }
\label{fig:flux}
\end{figure*}

\subsection{Individual distances: Single stars} \label{sec:singles}

Source positions at individual epochs were extracted by performing two-dimensional Gaussian fits with the AIPS task JMFIT. These and the associated uncertainties provided by JMFIT, which are based on the expected theoretical astrometric precision of an interferometer \citep{Condon_1997}, are listed in Table \ref{tab:positions}. We analyze the motion of all objects detected in at least 2 epochs. A total of 20 objects, which do not have a firm classification in the literature, show a motion consistent with that expected for background sources, i.e., their positions remain systematically unchanged within the positional errors, or even if they move, their derived parallaxes correspond to distances larger than 1 kpc.  This can be seen grapically in  Figure \ref{fig:flux}.  The horizontal axis of this plot corresponds to the {\it position change rate} in milli-arcseconds (mas) per year, which we define as the shift in position between consecutive epochs, normalized to one year, and averaged over all consecutive pairs of epochs.  The 20 unclassified  objects have position change rates below 3~mas~yr$^{-1}$, while objects that belong to Serpens or W40 clearly show larger values because of the significant contribution of their parallax and proper motion. We identify  these 20 objects  as background sources and give a ``B'' flag in Column 3 of Table \ref{tab:sources}. Note that not all these sources are necessarily extragalactic. Some might be Galactic objects located behind the  Serpens/Aquila complex. For example, the fit to the positions of the source PMN~1829+0101 yields a distance of $4.025^{+0.854}_{-0.600}$~kpc (Section  \ref{ref:pmn}). 
The large number of background sources detected here with the VLBA is not surprising. \cite{Oliveira_2009} determined that $25\%$ of the YSO candidates with IR excess in the Serpens/Aquila complex are actually background giants. As stated by  these authors, this is consistent with the location of the regions being close to the Galactic plane. 


Only 8 VLBA-detected objects are previously known YSOs, and one more object is a B1V star. Out of these 9 objects, two are resolved into double components in the GOBELINS data, while 7 are single stars. This gives a total of 11 individual objects. The astrometry of  5 single stars is given in the present section;  the other two single objects will be presented in a later paper because they were not detected often enough to do astrometric fits. 
The two binaries are  discussed in Section \ref{sec:multiple}.

\begin{figure*}[!ht]
\begin{center}
{\includegraphics[width=0.30\textwidth,angle=0]{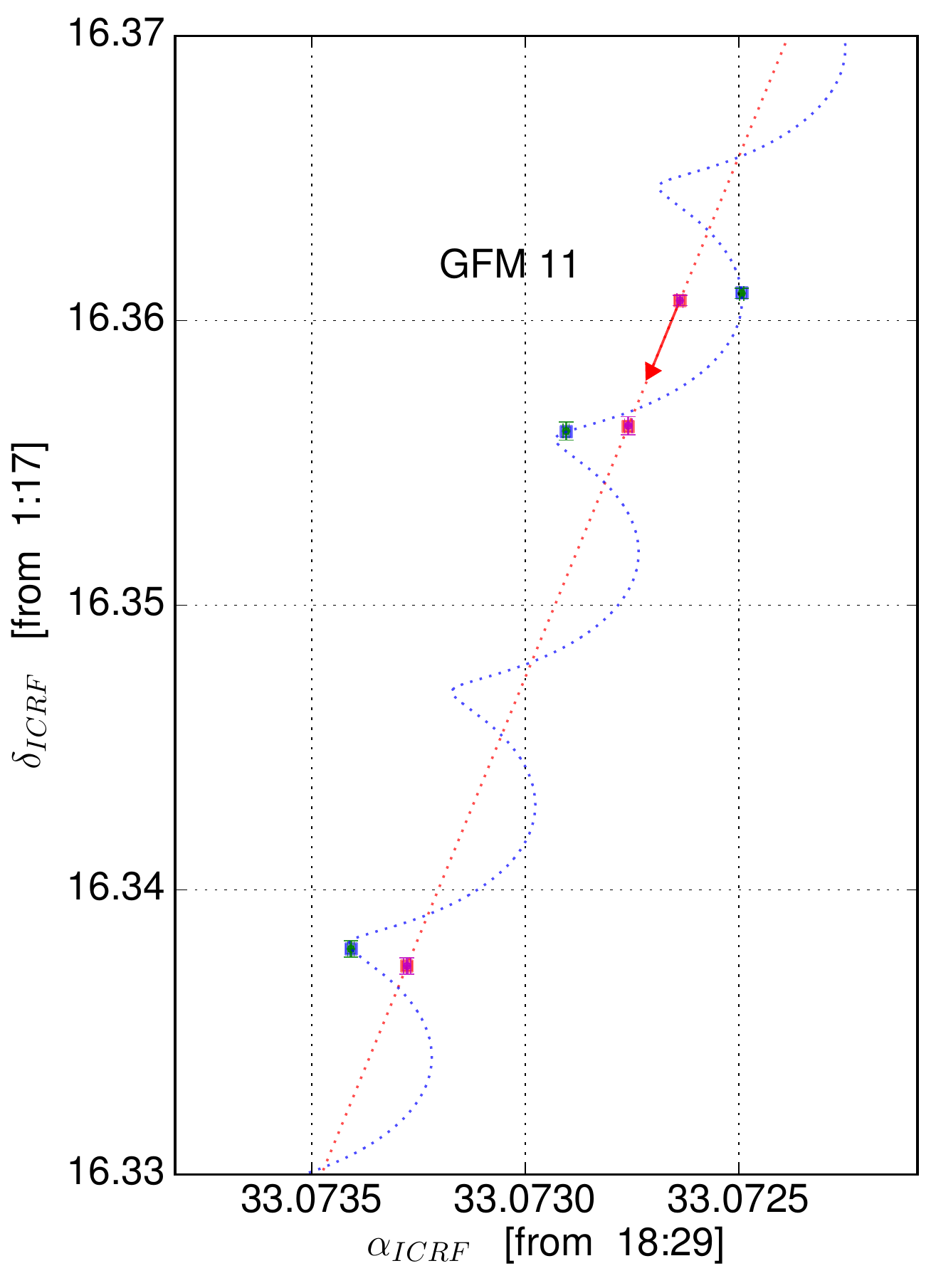}} 
{\includegraphics[width=0.224\textwidth,angle=0]{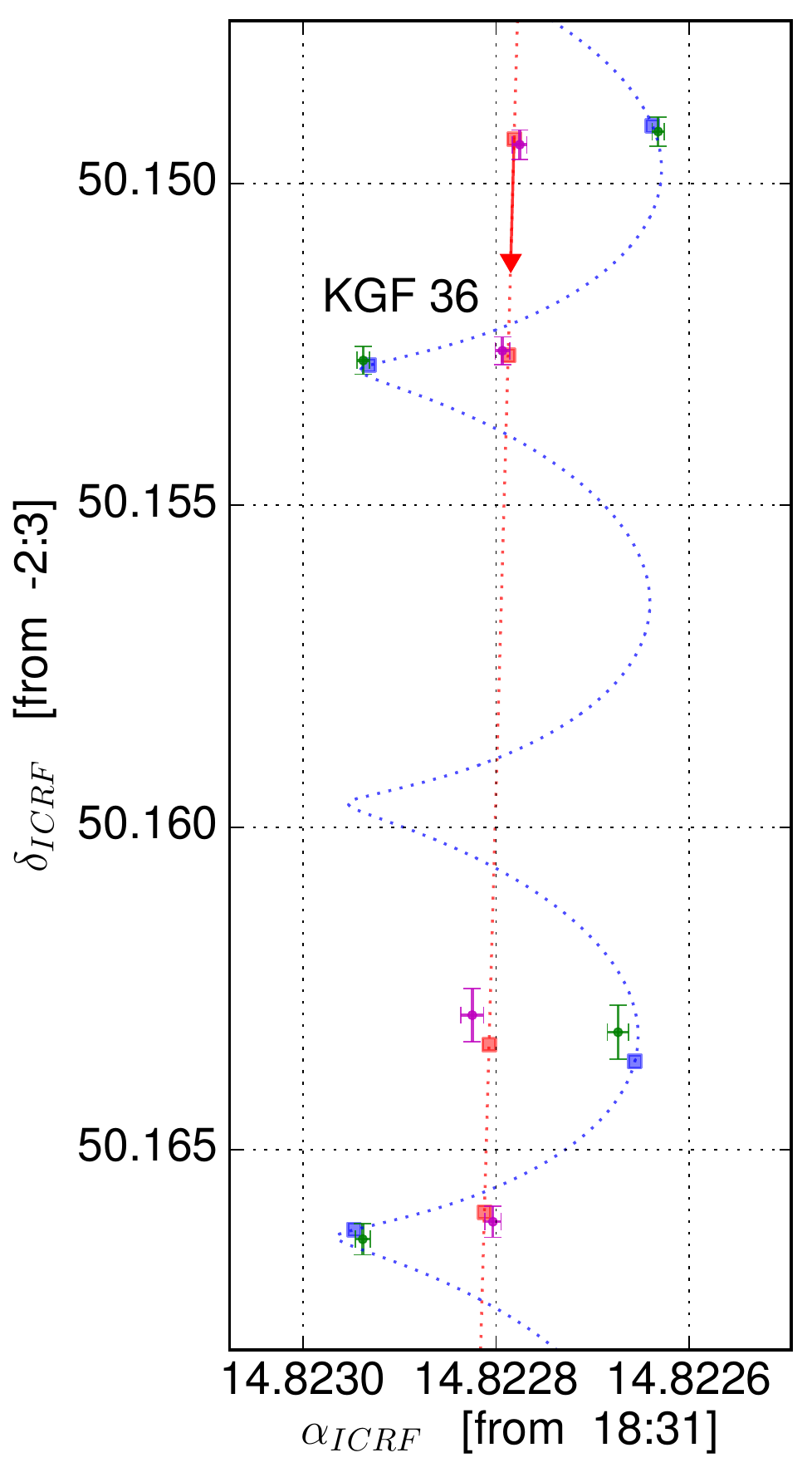}} 
{\includegraphics[width=0.28\textwidth,angle=0]{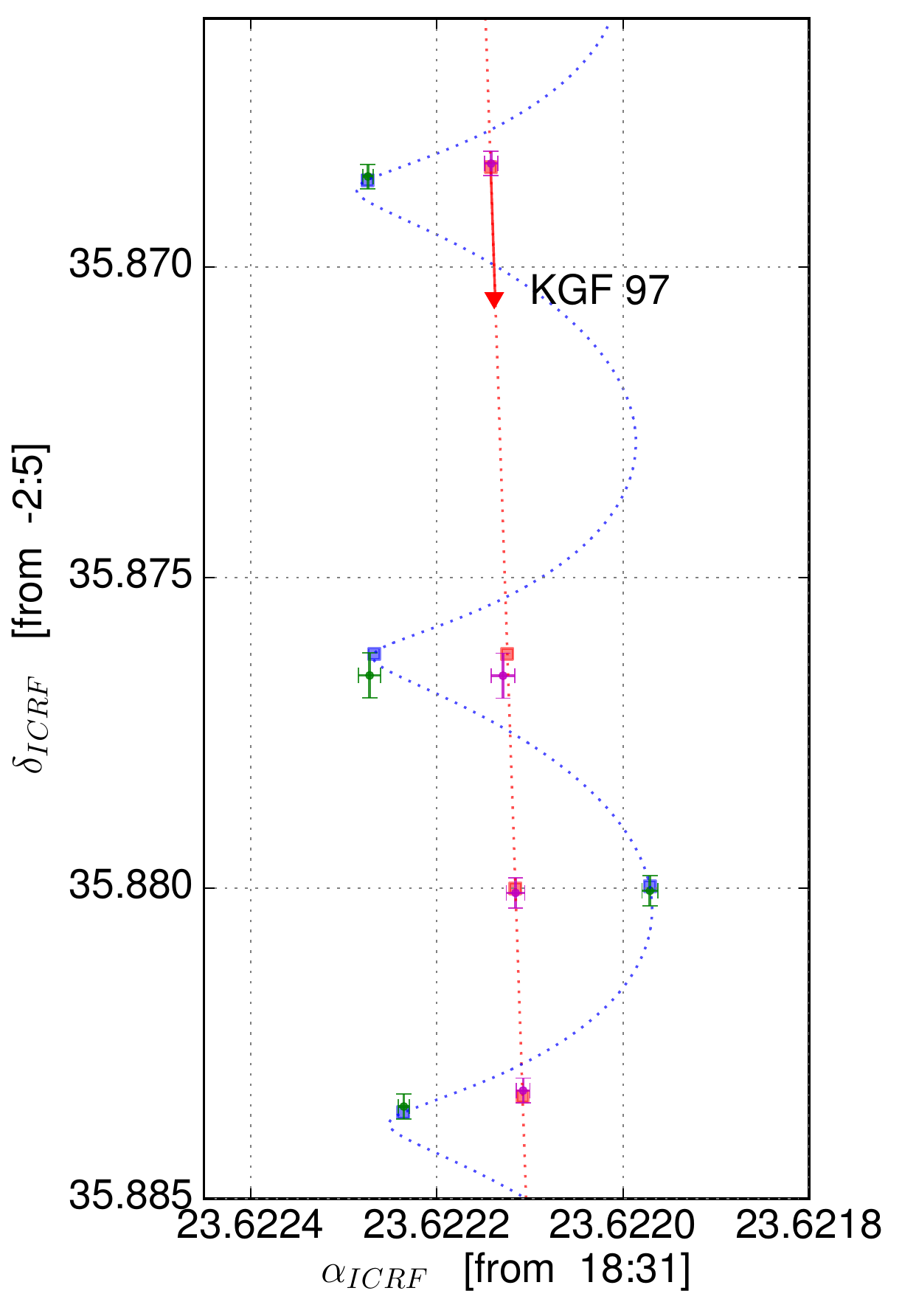}} \\
{\includegraphics[width=0.32\textwidth,angle=0]{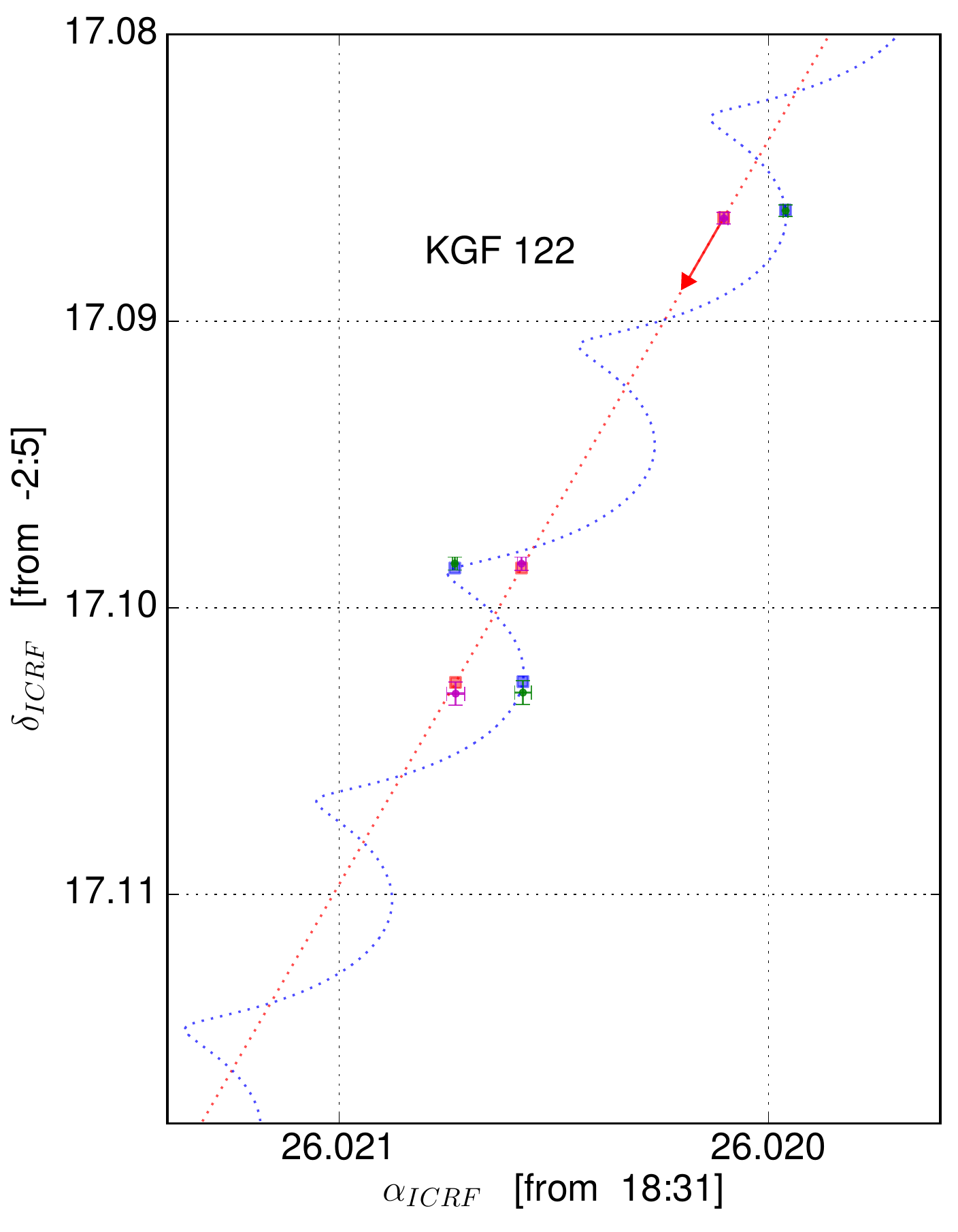}}
{\includegraphics[width=0.262\textwidth,angle=0]{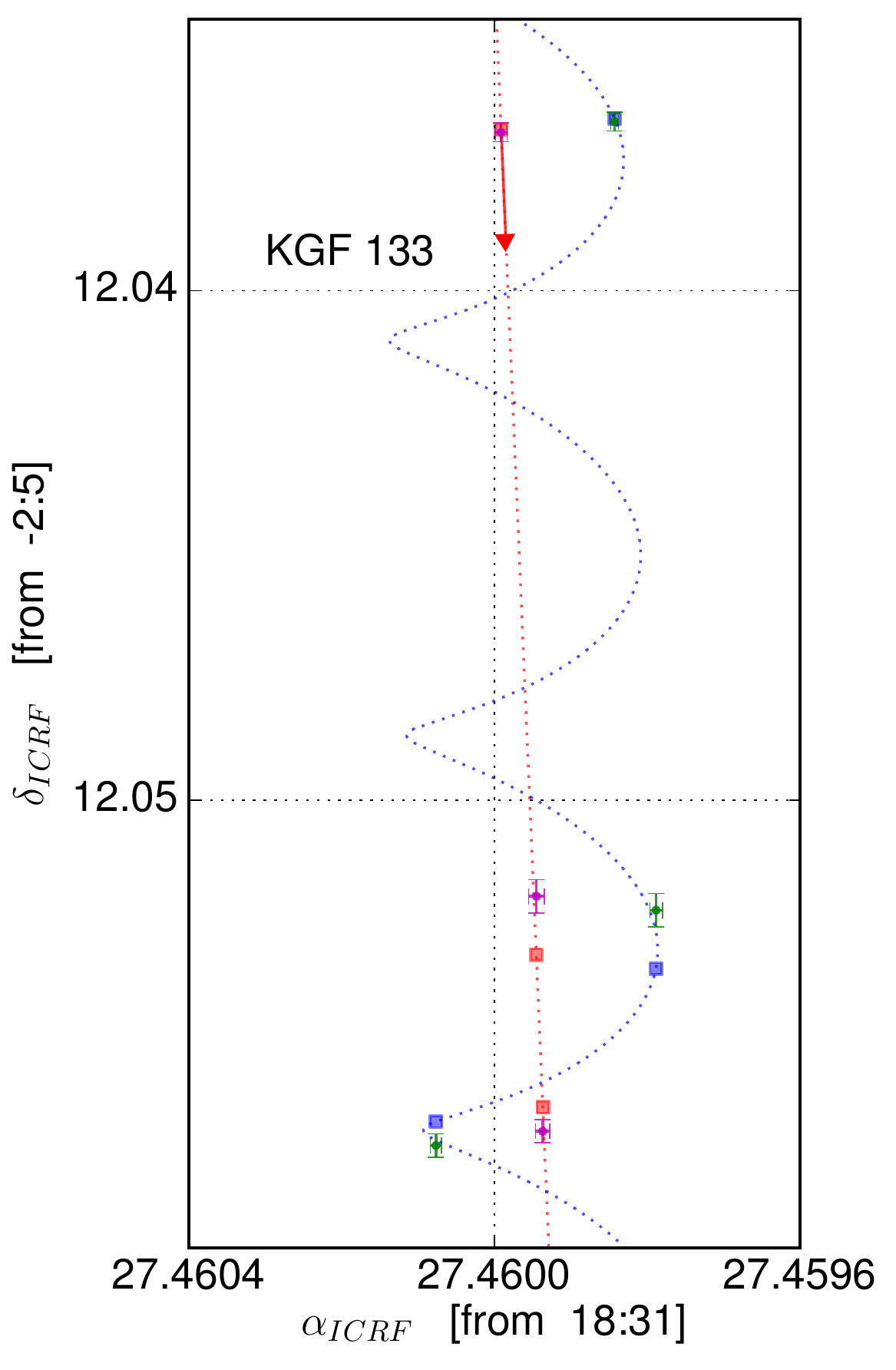}}
{\includegraphics[width=0.252\textwidth,angle=0]{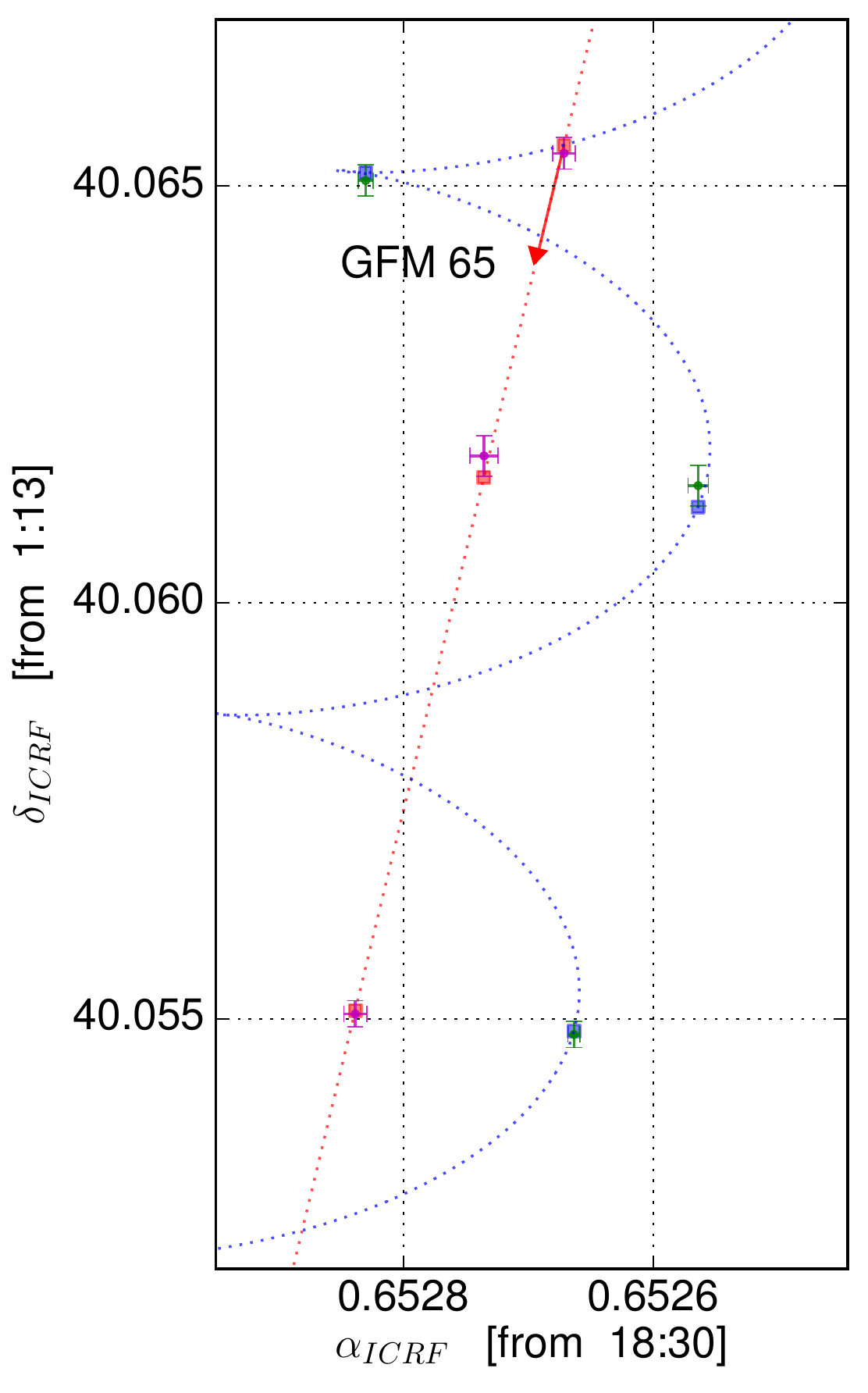}}
\caption{Observed positions and best fits for six sources. Measured positions are shown as green dots, and expected positions
from the fits as blue squares. The blue dotted line is the full model, and the red line is the model with the parallax signature removed. The
red squares indicate the position of the source expected from the model without parallax, while magenta dots are measured positions with parallax signature removed. The arrow indicates the direction of position change with time.
}
\label{fig:fit}
\end{center}
\end{figure*}

Parallax, $\varpi$, position at median epoch, $(\alpha_0,\delta_0)$, and proper motions $\mu_\alpha$, and $\mu_\delta$ are derived 
by fitting the equations
\begin{equation}
\alpha(t)= \alpha_0 + (\mu_\alpha\cos\delta)t + \varpi_\alpha f_\alpha(t),
\end{equation}
\begin{equation}
\delta(t)= \delta_0 +  \mu_\delta t + \varpi_\delta f_\delta(t),
\end{equation}

\noindent to  the measured positions and minimizing separately  $\chi_\alpha^2$ and $\chi_\delta^2$ along the right ascension and declination directions, respectively.  Here, $f_\alpha$ and $f_\delta$ are the projections of the parallactic ellipse over $\alpha$ and $\delta$, respectively.  The values of the parallax determined in right ascension ($\varpi_\alpha$) and declination ($\varpi_\delta$) were then weighted-averaged to produce a single parallax value. The fit is then repeated to solve for the remaining parameters while holding the best fit parallax solution constant. We show the resulting best fits 
in Figure \ref{fig:fit}, and summarize the derived astrometric parameters in Table \ref{tab:parallaxes}. Errors in the model parameters depend on the positional uncertainties of all the individual detections as measured by JMFIT.  However, systematic offsets in positions could be introduced by errors in station coordinates, Earth rotation parameters, reference source coordinates, and tropospheric zenith delays \citep{Pradel_2006}. When data from many epochs are available, these systematic offsets can be estimated by scaling the positional errors provided by JMFIT until the reduced $\chi^2$ of the fit becomes equal to 1 \citep[e.g.,][]{Menten_2007}. Here we are not able to apply this approach given that  we have typically 3--4 epochs available for each source.  We thus estimate systematic errors by using the empirical relations found by \cite{Pradel_2006}, according to which the VLBA astrometric accuracy scales linearly with the target to reference source angular separation. We obtain $\Delta\alpha\cos\delta=0.052-0.070$~mas and $\Delta\delta=0.124-0.182$~mas by extrapolating the astrometric errors given in Tables 3 and 4 in \cite{Pradel_2006}  for a source at a declination of $0^{\rm o}$ (the range in errors corresponds to the different source to calibrator angular separations). In order to estimate the offsets introduced by ionospheric phase delays, we follow the approach outlined  in Kounkel et. al.\ (2016, submitted). Source positions were referenced to a secondary phase calibrator by adding offsets such that the position of this secondary calibrator remains fixed in all epochs. We repeat the astrometric fits to the re-referenced target positions obtaining a different solution to that derived when all positions are referenced to  the main phase calibrator.  We take the difference in the distance solutions divided by the angular separation between the two phase calibrators as the phase gradient across the sky introduced by ionospheric delays. On average, this yields additional systematic offsets of $\Delta\alpha\cos\delta = 0.026$~mas and $\Delta\delta= 0.042$~mas in declination. In total, systematic errors of  $\Delta\alpha\cos\delta = 0.058-0.075$~mas and $\Delta\delta= 0.130-0.187$~mas were added quadratically to the statistical errors provided by JMFIT at each individual epoch and used in the last iteration of the  fits. 
  
 We discuss separately the properties of these objects in the following sections.  Sources names come from the X-ray surveys by \citet[][GFM]{Giardino_2007} and \citet[][KGF]{Kuhn_2010}. 

\subsubsection{GFM 11 = GBS-VLA J182933.07+011716.3}\label{sec:gfm11}

GFM 11 is a Class III YSO \citep{Giardino_2007}. Its spectral type remains somewhat uncertain: between G2.5 \citep{Winston_2010} and K0 \citep{Erickson_2015}.
The source has a spectral index\footnote{From VLA measurements published in  \cite{Ortiz_15}. The spectral index was taken between 4.5 and 7.5 GHz.} of $+0.3\pm0.2$, and shows high levels of variability  in both VLA \citep[$>73\%$;][]{Ortiz_15} and VLBA observations.  Based on optical spectroscopy, \cite{Erickson_2015} estimated a mass of $2.0~M_\odot$ for the source. 

\subsubsection{KGF 36 = GBS-VLA J183114.82-020350.1}

This source, identified as a main sequence star of B1 spectral type by \cite{Shuping_2012}, is located in the W40 cluster. Its radio flux as measured by the VLA shows variations of $44\pm9 \%$ on time scales of months at 4.5 GHz, and it has a spectral index of  $+0.3\pm0.2$. \cite{Shuping_2012} also suggested that KGF~36 is probably a binary source due to the presence of strong He I 1.083~$\mu$m absorption in the star spectra.  However, our VLBA observations have detected a single source with no sign of a close companion in the parallax fit.
Non-thermal  emission has been confirmed in other early-type B stars. The source S1 in Ophiuchus \citep{Andre_1988}  is perhaps the most documented case. \cite{Kuhn_2010} derived a photometric mass of 
$17~M_\odot$  from a color--magnitude J vs.\ J--H diagram assuming distance  of 600 pc and age of 1 Myr.

 
 \subsubsection{KGF 97 = GBS-VLA  J183123.62-020535.8}
 
KGF 97, whose spectral type is unknown, is also a YSO also located in the W40 cluster. Since the source does not show excess in the infrared $K_s$-Band, it is classified as a Class III object, with a mass of $3.3\pm1.0~M_\odot$ \citep[][reduced by a factor of $\sim2$ given a distance of 436~pc]{Kuhn_2010}. The source is found to be very variable in our VLBA observations by a factor $>10$. Additionally, it is one of the few sources of the cluster detected in circular polarization \citep{Ortiz_15}, a strong signature of gyrosynchrotron radiation.  The spectral index is $-0.1\pm0.1$.
 
\subsubsection{KGF 122 = GBS-VLA  J183126.02-020517.0 }

This source was classified as a low-mass Class II YSO by \cite{Shuping_2012} based on the analysis of infrared data.  It shows high flux variations in both VLA ($52\pm5 \%$ at 4.5 GHz) and VLBA observations, and has a negative spectral index of $-0.6\pm0.2$. \cite{Kuhn_2010} estimated a photometric mass of  $16~M_\odot$ 
for the source, and a bolometric luminosity of $2.9\times10^{4} L_\odot$, assuming 600~pc as the distance to the cluster (a lower distance of 436~pc reduces the luminosity and mass by a factor of $\sim2$).
Thus, the source may be associated to an early-type source.  We discard the last measured source position for the derivation of the astrometric  parameters because it significantly deteriorates the quality of the fit and, since we ignore the source of any positional error that may be introduced in this particular epoch, we cannot correct the  source position. 
  
\subsubsection{KGF 133 = GBS-VLA  J183127.45-020512.0}\label{sec:kgf133}

KGF 133 was identified as a Class II/III YSO by \cite{Mallick_2013} based on {\it Spitzer} and near-IR data. Like the rest of the VLBA-detected YSOs,  the source is very variable in radio, with fluctuations of $96\pm1 \%$ at 4.5 GHz \citep{Ortiz_15}. The spectral index of the source  is $+0.3\pm0.2$.  The mass of the source is not yet well constrained. \cite{Kuhn_2010} derived a photometric mass of $24~M_\odot$ (reduced to $\sim 12~M_\odot$ for a distance of 436~pc), but the associated error is uncertain and not provided by these authors. 
 
 \bigskip
 
 
\subsection{Multiple systems}\label{sec:multiple}

\subsubsection{GFM 65 = GBS-VLA J183000.65+011340.0} \label{sec:gfm65}

This source is a M0.5, $0.96~M_\odot$ star \citep{Winston_2010} located in the Serpens Core. It was classified as a Class III object by 
\citet{Giardino_2007}.
Based on multi-epoch VLA observations,  \cite{Ortiz_15} found that the source shows large flux variations ($>75\%$) on time scales of months,  and measured a spectral index of $-0.9\pm0.4$. Both properties of the radio emission are fully consistent with its non-thermal nature.  Because of this variability, the source has been detected with the VLBA just in 3 of the 6 observed epochs. Another source, possible a gravitationally bound companion, was detecetd
in two epochs separated by $\sim$5~mas from the primary. We are not able to constrain the orbit of the system using our present small number of detections. We perform the parallax fit for only one component following the procedure described in Section \ref{sec:singles}. 

\subsubsection{EC 95 = GBS-VLA J182957.89+011246.0}\label{sec:ec95}

EC 95 is located in the Serpens core. The system is formed by two close components first observed by \cite{Dzib_2010}. Early estimations of its spectral type ($\sim$K2 star), age ($\sim10^5$~yr) and mass ($\sim$4~M$_\odot$) indicated that the source is a proto-Herbig AeBe star \citep{Preibisch_1999}. \cite{Dzib_2011} reported observations from 11 epochs  taken with the VLBA at 8 GHz and reported a distance to the source of $429\pm2$~pc. Earlier, \cite{Dzib_2010} performed a circular Keplerian orbit fit to the data from 8 of these 11 epochs, constraining the orbital period of the system to $10-20$ years.  

\begin{figure*}[!ht]
\begin{center}
 \includegraphics[width=0.8\textwidth,angle=0]{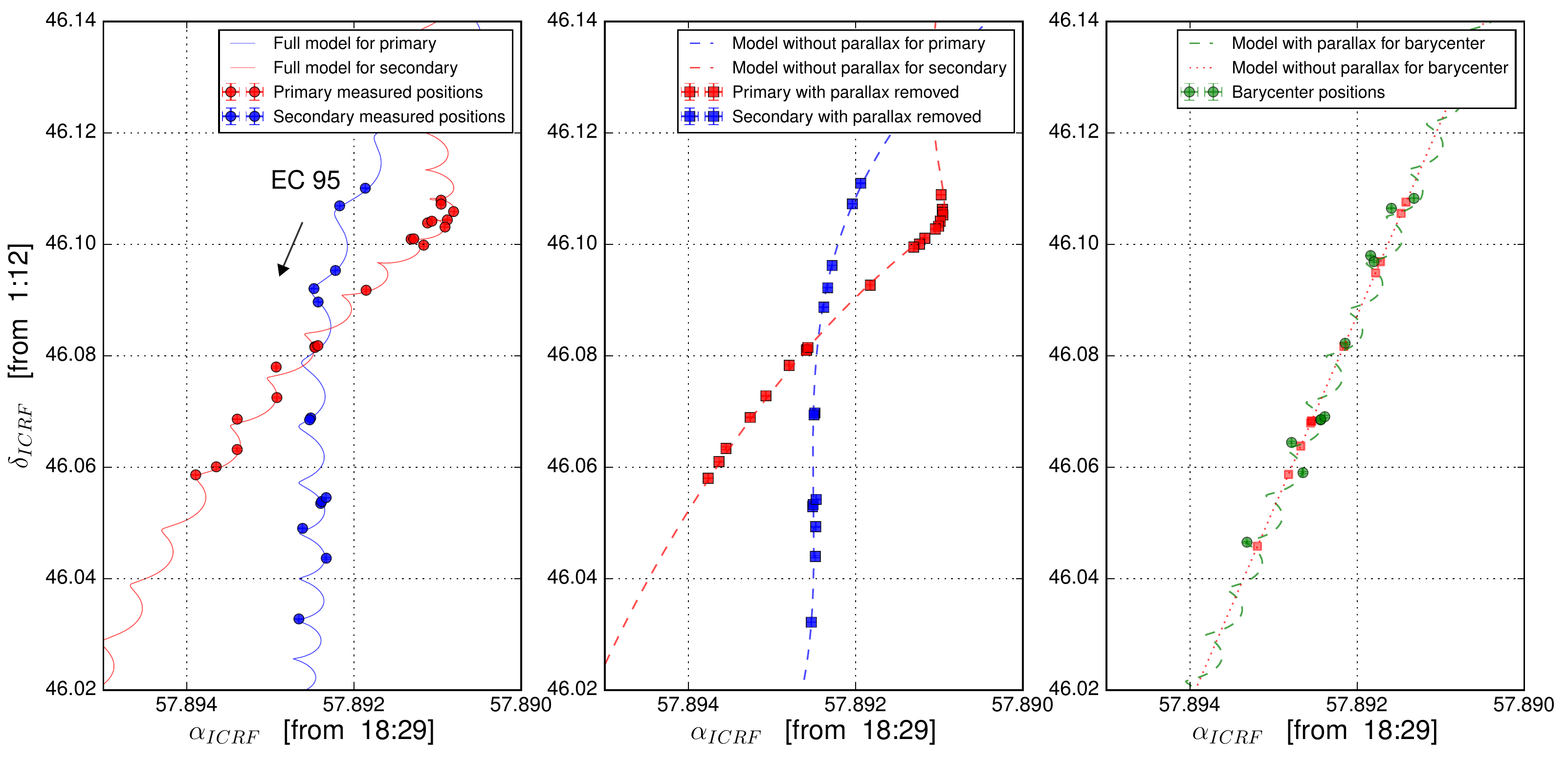}
\end{center}
 \caption{Observed positions of EC 95 and best astrometric fits.
{\it Left:} Measured positions of each component are shown as red and blue circles. The solid lines show the fit corresponding to the ``Full model'' described in the text.  The arrow indicates the direction of  position change with time. {\it Middle:} The squares mark the measured positions with the parallax signature removed, while the dashed lines are the fits from the ``Full model'',  also without parallax. {\it Right:} Green dots mark the position of the center of mass derived using the solutions from the orbital model for the mass ratio.  The green dashed line is the model for the motion of the center of mass of the system, while the red line is this same model with the parallax signature removed.  The red squares indicate the position of the center of mass expected from the model without parallax. 
}
\label{fig:ec95full}
\end{figure*}

\begin{figure*}[!ht]
\begin{center}
 \includegraphics[width=0.6\textwidth,angle=0]{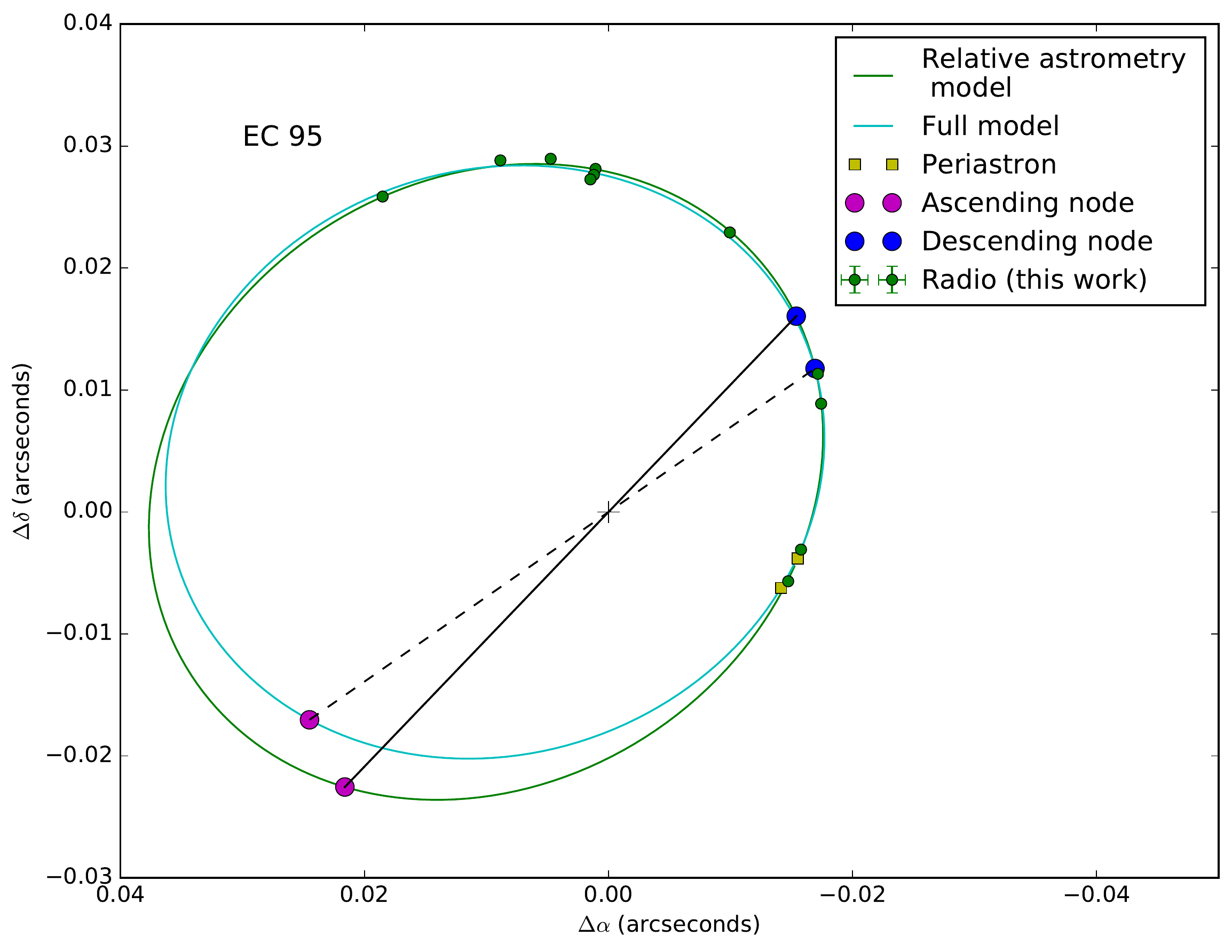}
\end{center}
 \caption{Relative positions of the components of the young binary system EC~95. The green points mark the detections with the VLBA.  Green and cyan solid lines correspond to the ``Relative astrometry" and ``Full model'' orbital fits, respectively. The black  solid and dashed lines trace the line of nodes of the  ``Relative astrometry" and ``Full model'', respectively.}
\label{fig:ec95orbit}
\end{figure*}

In order to derive the full orbital parameters of EC~95, we carried out follow-up VLBA observations as part of the project coded BD155, which observed the system at 8 GHz in 5 new epochs. 
The source has also been monitored with GOBELINS at 5 GHz, and  6 additional epochs are available. The new observations together with those previously reported by \cite{Dzib_2011} cover a baseline timescale of $\sim 8$ years, i.e.\ a significant fraction of the orbit.  Old data were recalibrated homogeneously applying the same calibration strategy as for the new data, and combined with the GOBELINS observations to form a single data set. The data were fitted with two models. In the first  ``Full model'', we fit the orbital and astrometric parameters of the system simultaneously. Orbital elements in this model are period ($P$), time of periastron passage ($T$), eccentricity ($e$), angle of line of nodes ($\Omega$), inclination ($i$), angle from node to periastron ($\omega$), semimajor axis ($a_1$) of the primary, and mass ratio ($m_2/m_1$). Astrometric parameters include center of mass at first  epoch of the GOBELINS observations where the primary is detected ($\alpha_{\rm{CM},0}$, $\delta_{\rm{CM},0}$), parallax ($\varpi$), and  proper motion ($\mu_\alpha$, $\mu_\delta$) of the system. For this fit, a grid of initial guesses of $P$, $e$, $T$, and $\omega$ is explored. The final values of these parameters are fine-tuned by the code, and the remaining model parameters are fitted directly.
The first panel of Figure \ref{fig:ec95full} shows the resulting best-fit curve and the measured positions of both components of the system, while the second panel shows the same fit without parallax and measured positions with parallax signature removed. Finally, the motion of the barycenter is shown in the last panel of the same figure. 

In the second ``Relative Model'', we use the {\it Binary Star Combined Solution Package} to fit the positions of the secondary relative to the primary component and solve for $P$, $T$, $e$, $\Omega$, $i$, $\omega$, and $a$. The total mass of the system is then derived from Kepler's law. The solutions found by the ``Full model'' are used as initial guesses for this fit. Uncertainties in the orbital elements are computed from the scatter on model parameters. The best-fit solution is shown in Figure \ref{fig:ec95orbit}, and compared with the solution found by the ``Full model''. Solutions for the orbital elements from both models are given in Table \ref{tab:ec95orbit_par}.  \cite{Dzib_2010} argued that one of the system components should be considerably more massive than the other, however, their reported observations only covered a small fraction of the complete orbit. Here, based on a larger number of observations, we have derived a similar mass for both components, while the total mass of $\sim 4 M_\odot$ is consistent with that estimated in past works \citep{Preibisch_1999,Pontoppidan_2004}, and with the spectral type of the source.

\begin{figure*}[!ht]
\begin{center}
 \includegraphics[width=1.0\textwidth,angle=0]{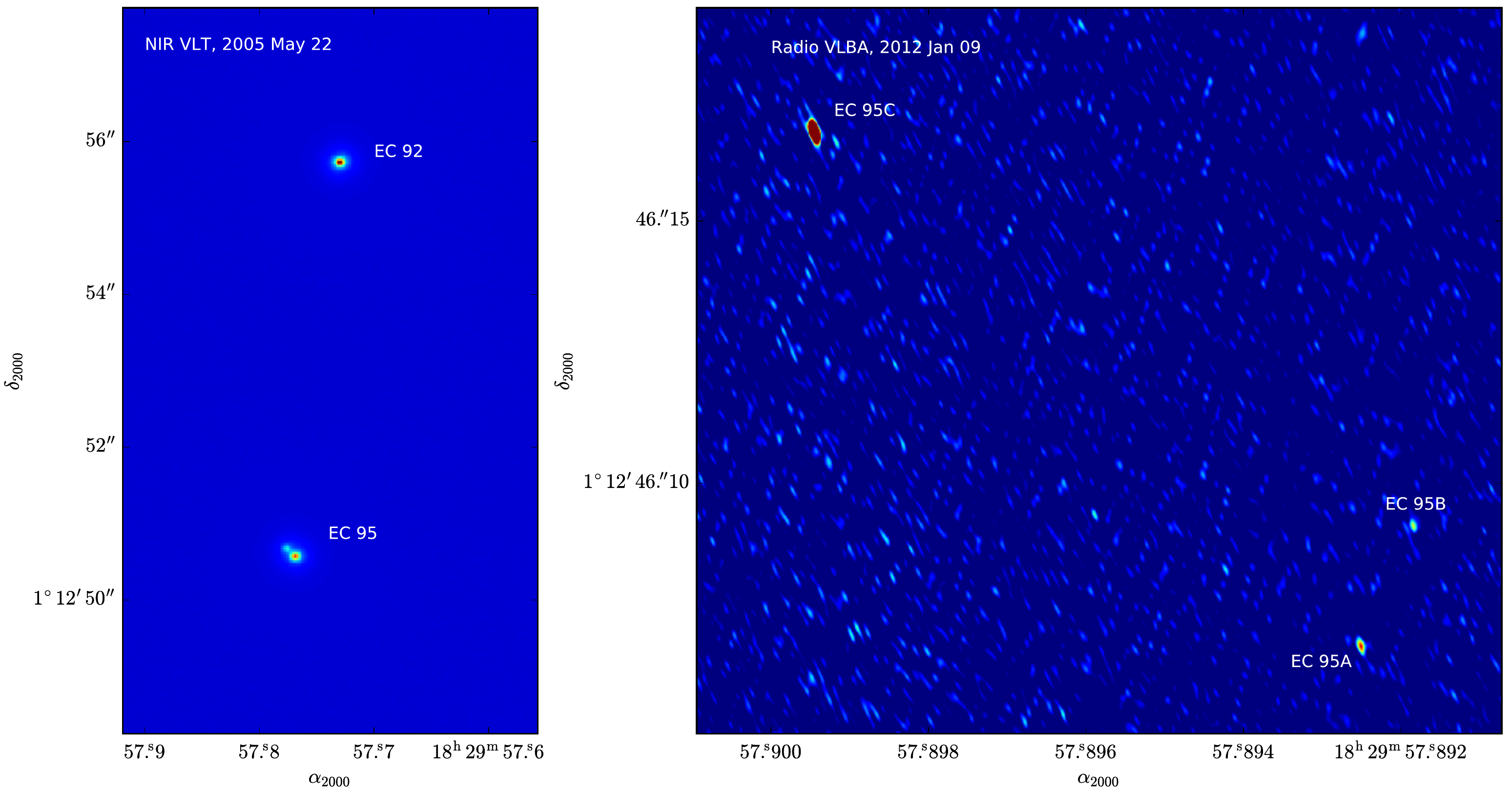}
\end{center}
 \caption{{\it Left}: NIR image of EC~92 and EC~95 taken with the VLT.  {\it Right}: Radio image of the system EC~95 corresponding to one of the two epochs when the three sources are  detected simultaneously with the VLBA.  }
\label{fig:ec95nir}
\end{figure*}

A third source  is detected in the EC95 system at two epochs
at 8 and 61$\sigma$, respectively. This source was located $\sim145$~mas to the North-east of the barycenter of the close binary, 
 at a position angle of $48.3^{\rm o}$ on 2008 September 15, and $\sim138$~mas in the same direction,  at a position angle of $52.4^{\rm o}$ on  2012 January 9 (See Table \ref{tab:ec95}).  The third source was also detected in near-IR (NIR) observations taken at the VLT  on 2005, May 11 and 22 \citep{Duchene_2007}.  A map of the system as seen in NIR emission is shown in the first panel of Figure \ref{fig:ec95nir}. The northernmost source is EC~92, a young ($\sim 10^5$~yr), Class I \citep{Pontoppidan_2004} and low-mass  \citep[$\sim 0.5~M_\odot$;][]{Preibisch_1999} star. EC~95 is the brightest source to the south in the map. While the two close components of EC~95 are unresolved, the third component is clearly visible, at a position angle of $47.2^{\rm o}$, and separation of $152~{\rm mas}$ from the close binary, i.e.\ at a position similar to that of the radio source seen in our VLBA images (Figure \ref{fig:ec95nir}, right).  Given the short angular separation of the third component relative to the close binary, it is possible that the three sources form a bound system. To investigate this possibility, we include  two more free parameters in the  ``Full model'', corresponding to the acceleration terms in right ascension, $a_\alpha$, and declination, $a_\delta$. We find that these acceleration terms are zero within the errors, and that the motion of the barycenter of the close binary remains linear during the timescale covered by our observations. 
This suggests that the third source may be much less massive than the close binary and following a very long period orbit. Actually, if we assume that the total mass of the system is $4.2~M_\odot$, i.e.\ that the mass of the third source is negligible, we estimate an orbital period $\sim 260$~yr.  The change in angular separation of the third companion (detected first in the NIR and then in the VLBA images) relative to the barycenter of the close binary  is $\sim20$ mas in 6.7 yr, while the position angle only changes $\sim5$ degrees over this time scale. This is consistent, within the errors, with the expected motion of the companion on a circular orbit that has the period  estimated  above. Unfortunately, the third companion has remained undetectable in the radio since 2012. If there were more detections, we would constrain its astrometric parameters and  investigate a possible acceleration induced by its orbital motion around the close binary. 

Finally, we note that \cite{Dzib_2011} estimated a distance to EC~95 of $429\pm2$~pc by modeling separately the source motion of each component as a superposition of parallax and uniform accelerated proper motion. The derived distance from the ``Full model'' is $435.2\pm6.0$~pc, which is consistent within $1\sigma$ with the previous determination.

\subsection{Comments on other sources: PMN 1829+0101}\label{ref:pmn}


PMN 1829+0101 = GBS~VLA J182930.71+010048.3 is a strong radio source with reported VLA fluxes of $196.0\pm5.9$~mJy  at 1.4~GHz \citep{Ofek_2011} and $32.10\pm5.60$~mJy at 4.5 GHz \citep{Ortiz_15}. The source shows an extended structure of $\sim 1.4''$ in the 4.5~GHz VLA images, but this emission is filtered out  by the VLBA. There is a counterpart  in  X-ray emission at $\sim1.3''$ \citep{Xmm_2013}, and in the IRAC 3.6~$\mu$m-band \citep{Evans_2003} at $\sim 1.7''$ from the radio peak. The fit to the data yields $\varpi=0.248\pm0.044$~mas, corresponding to a distance of  $d=4025^{+854}_{-600}$~pc. 
Using this distance we derive the location  $(x,y,z)$ of the source in the Milky Way. This position is expressed in the rectangular frame centered on the location of the Sun, with the ($Ox$) axis pointing toward the Galactic Center, the ($Oy$) axis perpendicular to ($Ox$) and pointing in the direction of the Galactic rotation, and ($Oz$)   pointing toward the Galactic North Pole.  The source coordinates in this system are $(x,y,z) = (3443, 2096, 377)$~pc: it is located in the direction of the Scutum arm, which hosts newly formed OB-type stars, but at 377~pc  above the Galactic mid-plane. 

\section{Discussion}\label{sec:discussion}

\begin{figure}[!ht]
\begin{center}
 \includegraphics[width=0.3\textwidth,angle=0]{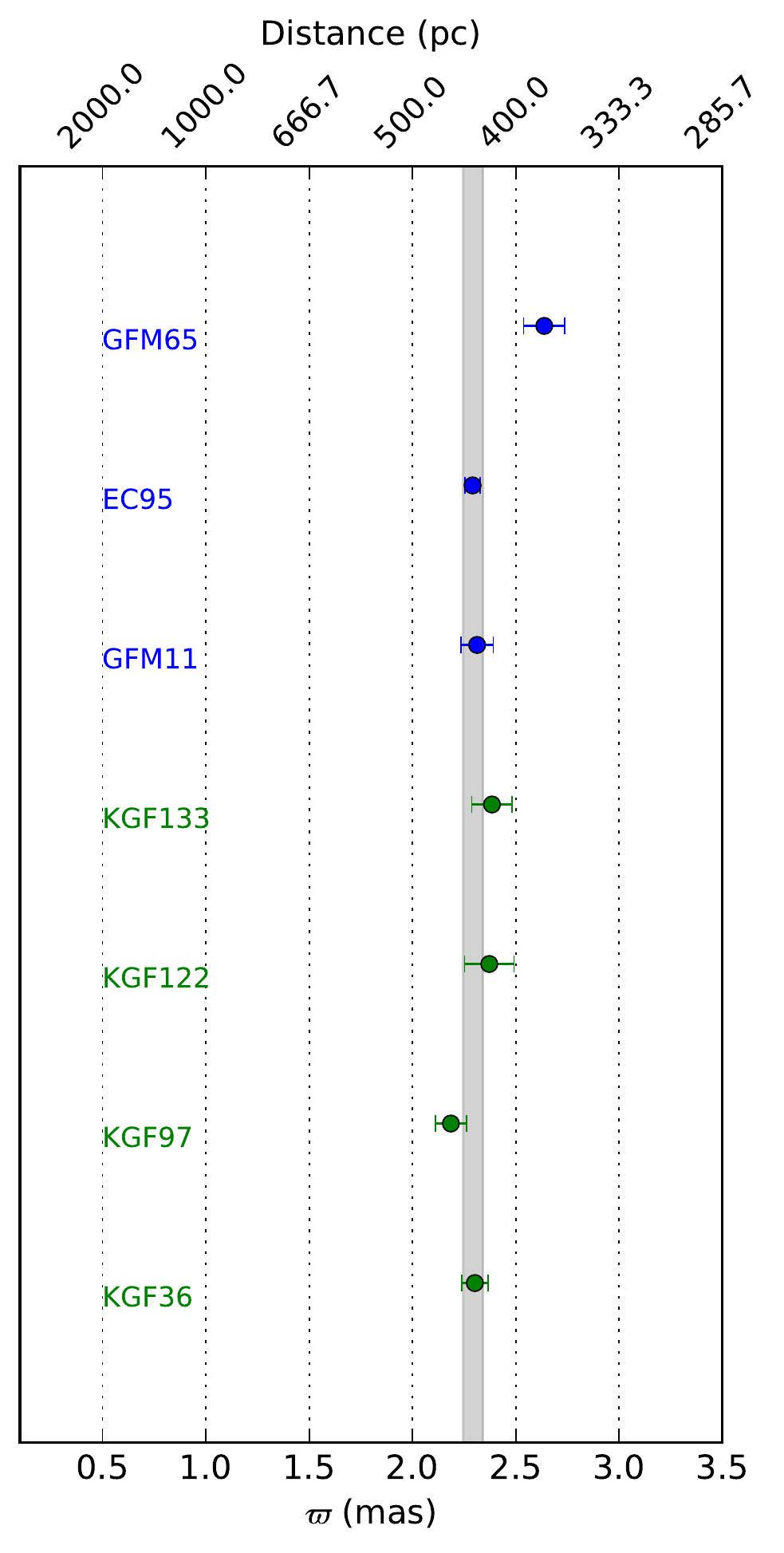}
 \includegraphics[width=0.45\textwidth,angle=0]{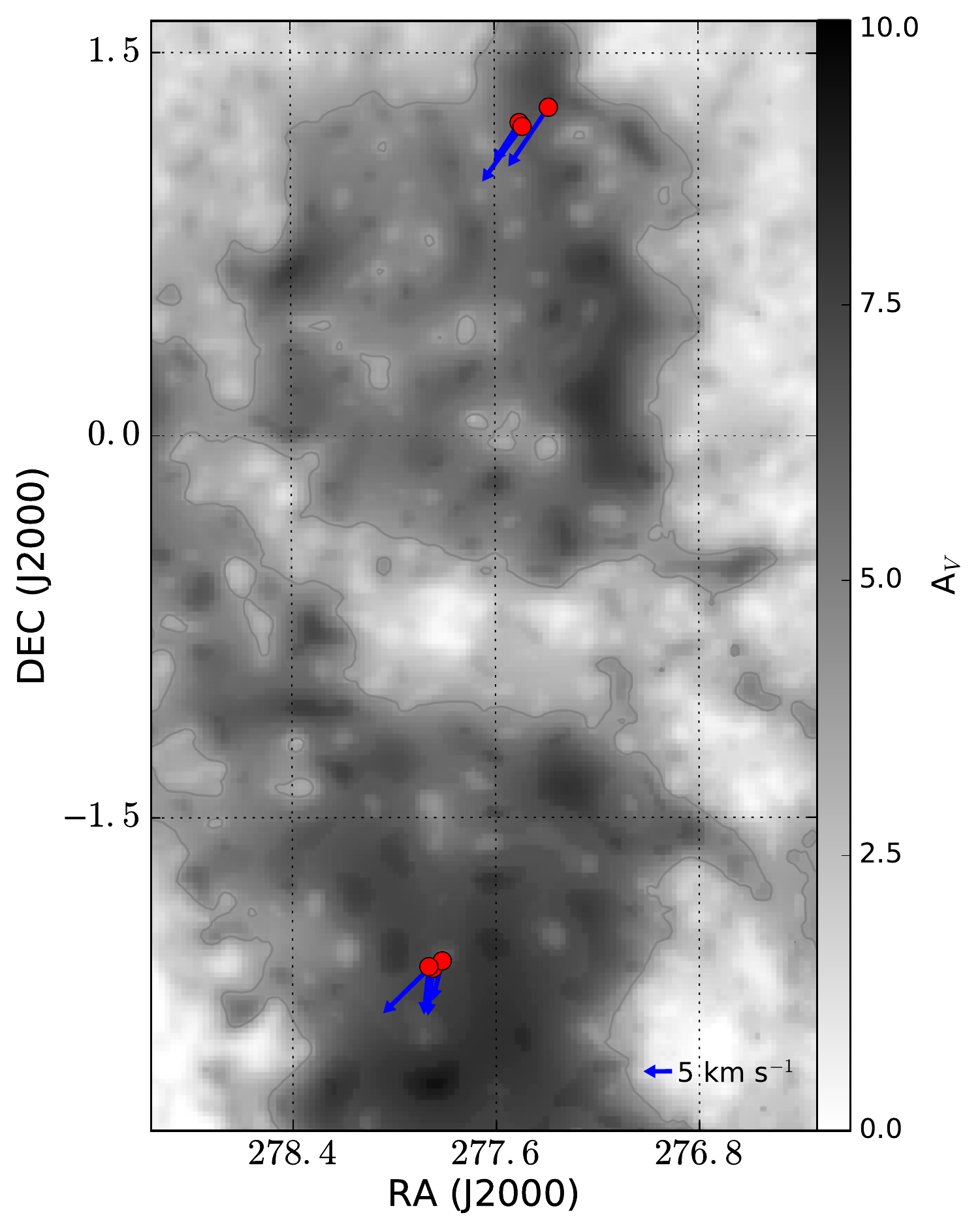}
\end{center}
 \caption{{\it Left:} Parallax results for the Serpens/Aquila complex. Blue circles and characters are for sources in the Serpens Core, while green circles and characters are for sources W40. The grey vertical bar shows the mean parallax value for all sources but GFM~65 (a binary source for which we are not able to model its orbital motion based on the data collected so far), and its standard deviation (see text). {\it Right:}  Enlargement of Figure \ref{fig:ser_aq} showing the spatial distribution of the YSOs in Serpens and W40 with astrometric parameters derived in this work. The arrows represent the source tangential velocity corrected by the solar motion.}
\label{fig:prlx}
\end{figure}

We have derived the distance to 7 objects  in the Serpens/Aquila complex. The parallaxes for these objects  are shown graphically in Figure \ref{fig:prlx}, where we see clearly that sources in Serpens and Aquila share similar values. Proper motions, on the other hand, show a large spread, but this is expected as Serpens and W40 are different clusters.  The weighted mean value of the 7 parallaxes is $\varpi = 2.32$~mas, with a  weighted standard deviation of $\sigma_\varpi = 0.10$~mas. Only the source  GMF~65, for which we derive $379.1\pm17.0$~pc, differs from the rest by more than 1 sigma.  As discussed in Section \ref{sec:gfm65}, this source seems to be a binary system, whose orbital motion remains unmodelled because of the low number of detections. Ignoring this source yields a mean weighted parallax of $\varpi = 2.29\pm0.05$~mas. This corresponds to a weighted mean distance of $d = 436.0$~pc, with a standard deviation of $\sigma_d=9.2$~pc. The standard deviation on the mean reflects only the uncertainties in the distance measurements because typical errors on individual distances are larger than 10~pc. 

Note that \cite{Straizys_2003} determined  the near edge of the Aquila/Serpens cloud complex to be at $225\pm55$~pc, with a depth of 80~pc. Therefore, according to their estimates, the far edge of the complex lies at a distance 
(assuming a $+1\sigma$ deviation)  of $225+55+80=360$~pc \citep{Winston_2010}. Assuming a $+3\sigma$ deviation, we place the far edge of the cloud at a maximum distance of 470~pc, which is consistent with the mean distance to the cloud obtained here from parallax measurements. 

Our measurements not only confirm the early estimation by \cite{Dzib_2010} of a larger distance to Serpens than previously thought, but also show that Serpens and W40 are part of the same complex, lying at the same distance along the line of sight. Earlier estimates based on indirect methods, e.g.\ by \cite{Kuhn_2010}  and \cite{Shuping_2012}, suggested a mean distance to W40 of $\sim500\pm50$~pc. Based on these measurements, W40 and the Serpens region have been treated in the literature as  separated objects, as such works used the value obtained by \cite{Straizys_1996} of $\sim$260~pc for the Serpens/Aquila Rift \citep[see e.g.][]{Straizys_2003,Gutermuth_2008}. When it was first discovered, Serpens South
was associated with Serpens Main because  both regions share  similar local standard of rest (LSR) velocities \citep[$\sim$6--11 km s$^{-1}$;][]{White_1995, Gutermuth_2008, Bontemps_2010}, indicating that they are comoving. Until recently, it has become more common to consider that Serpens South and W40 form a single cloud structure. This is because the LSR velocities  measured by molecular line observations in the entire W40/Serpens South region are 4--10~km~s$^{-1}$ \citep{Zeilik_1978, Maury_2011}, which are in the range of the LSR velocities measured in Serpens Main. 
We do not have any astrometric measurement to sources  in Serpens South \citep[because known YSOs in the cluster are intrinsically radio weak,][]{Ortiz_15,Kern_2016}, but given its proximity and similarity in LSR velocities to W40, we speculate that these two clusters are physically associated. If this last statement is confirmed, it would represent a meaningful  evidence for an association between Serpens Main, W40, and Serpens South.  

Proper motions are plotted in Figure \ref{fig:prlx} after the correction for  the solar peculiar motion is applied. Mean values are $(\mu_\alpha\cos\delta=8.0\pm2.2~{\rm mas~yr}^{-1},\mu_\delta = -11.6\pm2.9~{\rm mas~yr}^{-1})$  for Serpens sources, and $(\mu_\alpha\cos\delta=3.8\pm4.1~{\rm mas~yr}^{-1},\mu_\delta = -10.2\pm0.9~{\rm mas~yr}^{-1})$ for W40 sources. 	Uncertainties in the mean values correspond to the standard deviation of individual measurements in each cluster.  It appears that the clusters are moving in similar directions, which is an additional support for our interpretation of Serpens Main and W40 being part of the same cloud complex. 

Finally, we note that a larger distance to the Serpens and Aquila regions imply luminosities and dust masses larger by a factor of $\sim2.8$ relative to those derived assuming 260~pc, and makes young stellar objects younger with respect to evolutionary tracks.
 This implies that the physical interpretation of the stellar and core population in the regions needs to be revised. 




\section{Summary}\label{sec:summary}

We have analyzed multi-epoch VLBA observations taken as part of the GOBELINS project  toward young stars in the Serpens and W40 regions in the Aquila complex. The astrometric fits to 7 sources, including one confirmed binary (posible triple) system, provide us with parallaxes and proper motions for single sources, as well as with the orbital parameters for the multiple system. Since individual parallaxes of sources in Serpens are consistent with those of W40 sources, we conclude that both Serpens and W40 are located at the same distance. The mean parallax value yields $436.0\pm9.2$~pc, confirming the early distance estimation obtained solely for the source EC~95 in the Serpens Core, which was also derived from a parallax measurement with the VLBA. The other $20$ sources detected during the survey resulted to be background sources,  not associated with the Aquila Rift.  

\acknowledgements{G.N.O.-L., L.L., L.F.R., G.P., and J.L.R. acknowledge DGAPA, UNAM, CONACyT, Mexico for financial support. L.L. and G.N.O.-L. also acknowledge support from von Humboldt Stiftung. N.J.E. was supported by NSF grant AST-1109116 to the University of Texas at Austin. P.A.B.G. acknowledges financial support from FAPESP. The National Radio Astronomy Observatory is operated by Associated Universities, Inc., under cooperative agreement with the National Science Foundation.
}

\bibliographystyle{aasjournal}
\bibliography{serpens-submitted.bib}


\newpage

\clearpage
\begin{deluxetable*}{llllc}
\tabletypesize{\scriptsize}
\tablecaption{Observed epochs \label{tab:obs}}
\tablewidth{0pt}
\tablehead{ Project & Observation & \multicolumn{2}{c}{VLBA pointing positions} & Observed \\
code &  Date  & \colhead{R.A. ($\alpha_{2000}$)} &  \colhead{Decl. ($\delta_{2000}$)} & band \\
}\startdata
BL175E0 &01 sep 2013 & 18:29:10.178 & +01:25:59.56 & C\\
        &            & 18:29:27.366 & +01:20:37.43 & \\
BL175E1 &02 sep 2013 & 18:29:30.714 & +01:00:48.31 & C\\ 
        &            & 18:29:47.838 & +01:14:21.66 & \\
BL175E2 &03 sep 2013 & 18:30:44.115 & --02:01:45.66 & C\\
        &            & 18:31:21.969 & --02:04:52.54 & \\
BL175E3 &05 sep 2013 & 18:29:49.507 & +01:19:55.88 & C\\
        &            & 18:29:52.736 & --01:51:59.93 & \\
BL175E4 &07 sep 2013 & 18:31:21.141 & --02:04:31.08 & X\\
        &            & 18:29:16.120 & +01:04:37.58 \\
        &            & 18:29:33.073 & +01:17:16.39 \\
BL175DX & 17 feb 2014& 18:31:18.685 & --01:54:55.99 & X\\
BL175G0 &01 mar 2014 & 18:29:10.178 & +01:25:59.56 & C\\
        &            & 18:29:27.366 & +01:20:37.43 \\
BL175G1 &03 mar 2014 & 18:29:30.714 & +01:00:48.31 & C\\
        &            & 18:29:47.838 & +01:14:21.66 \\
BL175G2 &04 mar 2014 & 18:31:21.969 & --02:04:52.54 & C\\
        &            & 18:30:44.115 & --02:01:45.66 \\
BL175G3 &06 mar 2014 & 18:29:49.507 & +01:19:55.88 & C\\
        &            & 18:29:52.736 & --01:51:59.93 \\
BL175G4 &09 mar 2014 & 18:29:16.120 & +01:04:37.58 & X\\
        &            & 18:29:33.073 & +01:17:16.39 \\
BL175GC &01 apr 2014 & 18:28:54.46  & +01:18:23.78 & C\\
        &            & 18:29:48.83  & +01:06:47.46 \\
BL175CR &07 oct 2014 & 18:29:10.178 & +01:25:59.56& C\\
        &            & 18:29:27.366 & +01:20:37.43 \\
BL175CS &12 oct 2014 & 18:29:30.714 & +01:00:48.31& C\\
        &            & 18:29:47.838 & +01:14:21.66 \\
BL175CT &15 oct 2014 & 18:31:21.969 & --02:04:52.54 & C\\
        &            & 18:30:44.115 & --02:01:45.66\\
BL175EX & 27 feb 2015& 18:29:10.178 & +01:25:59.56& C\\
        &            & 18:29:27.366 & +01:20:37.43 \\
BL175EY & 02 mar 2015& 18:29:47.838 & +01:14:21.66  & C\\
        &            & 18:31:18.685 & --01:54:55.99   \\
BL175EZ & 20 mar 2015& 18:31:21.969 & --02:04:52.54 &C\\
        &            & 18:30:44.115 & --02:01:45.66  \\
BL175GT & 15 sep 2015& 18:28:54.46  & +01:18:23.78  &X\\
        &            & 18:29:48.83  & +01:06:47.46 \\
BL175GU & 19 sep 2015& 18:31:21.969 & --02:04:52.54 &C\\
        &            & 18:30:44.115 & --02:01:45.66  \\
BL175GW & 04 oct 2015& 18:29:10.178 & +01:25:59.56 & C\\
        &            & 18:29:27.366 & +01:20:37.43 \\
BL175GX & 06 oct 2015& 18:29:47.838 & +01:14:21.66 & C\\
        &            & 18:31:18.685 & --01:54:55.99   \\
BL175GV & 11 oct 2015& 18:31:21.141 & --02:04:31.08 &C\\
        &            & 18:29:16.120 & +01:04:37.58 \\
        &            & 18:29:33.073 & +01:17:16.39 \\
BL175GY & 13 oct 2015& 18:29:49.507 & +01:19:55.88   & C\\
        &            & 18:29:52.736 & --01:51:59.93  \\
BL175CU & 29 feb 2016& 18:29:52.736 & --01:51:59.93    & C\\
        &            & 18:31:21.141 & --02:04:31.08  \\
BL175F4 & 20 mar 2016& 18:29:33.073 & +01:17:16.39  & X\\
BL175F8 & 28 apr 2016& 18:29:47.838 & +01:14:21.66  & C\\
\enddata
\end{deluxetable*}
\clearpage

\floattable
\clearpage
\begin{deluxetable}{cccccccc}
\tabletypesize{\scriptsize}
\tablewidth{0pt}
\tablecolumns{6}
\tablecaption{Calibrators setup}
\tablehead{   R.A.       & DEC.   &  Calibrators$^1$ \\
                     (J2000)  & (J2000)  & \\ 
}
\startdata
  18:29:52.736  & --01:51:59.93   & \multirow{2}{*}{J1834--0301, J1833+0115, J1824+0119,   J1821--0502}  \\
  18:31:21.141  & --02:04:31.08 &  \\ \hline
  18:29:47.838  & +01:14:21.66 &  \multirow{2}{*}{J1833+0115, J1826+0149,  J1824+0119, J1832+0118} \\
  18:29:30.714  & +01:00:48.31 & \\ \hline
  18:28:54.460  & +01:18:23.78 &  \multirow{2}{*}{J1832+0118, J1833+0115, J1826+0149, J1824+0119} \\
  18:29:48.830  & +01:06:47.46  &  \\ \hline
  18:31:21.969  & --02:04:52.54  &  \multirow{2}{*}{J1834--0301, J1833+0115, J1824+0119, J1821--0502} \\
  18:30:44.115  & --02:01:45.66  & \\ \hline
  18:29:16.120  & +01:04:37.58 &  \multirow{5}{*}{J1826+0149, J1833+0115, J1824+0119, J1832+0118} \\
  18:29:33.073  & +01:17:16.39 & \\ 
  18:29:10.178  & +01:25:59.56 &  \\ 
  18:29:27.366  & +01:20:37.43 & \\ 
  18:29:33.073  & +01:17:16.39  & \\ \hline
  18:31:18.685  & --01:54:55.99  & J1834--0301, J1824+0119, J1819--0258, J1833+0115 \\ \hline 
  18:29:49.507  & +01:19:55.88 & J1826+0149, J1832+0118, J1833+0115,  J1824+0119 \\
 %
\enddata
\tablenotetext{1}{First source in the list corresponds to main phase calibrator. }
\label{tab:point}
\end{deluxetable}
\clearpage

\floattable
\clearpage
\begin{deluxetable}{cccccccccc}
\tabletypesize{\scriptsize}
\tablecaption{Detected sources \label{tab:sources}}
\tablewidth{10pt}
\tablehead{ GBS-VLA&       Other & Type of & Minimum flux   & Maximum flux  & Minimum flux  & Maximum flux   & log [$T_b$ (K) ]$^3$ & SED             \\
        name$^{1}$      &  identifier &    source  & at 5 GHz    &at 5 GHz & at 8 GHz  & at 8 GHz   &                  &    Class       \\
                                &                 &             &     (mJy)       &  (mJy)      &  (mJy) &  (mJy)        &                        \\
                  (1)          &         (2)   &   (3)     &  (4)          &   (5)         &    (6)         &     (7)          &  (8)      &  (9)           \\
}\startdata
\sidehead{Serpens Main}
\hline
J182854.44+011859.7 & -- & ? & \multicolumn{2}{c}{--} & \multicolumn{2}{c}{0.26$\pm$0.05}  &  $>$6.4 & --  \\
J182854.46+011823.7 & -- & B & \multicolumn{2}{c}{3.88$\pm$0.05} & \multicolumn{2}{c}{6.04$\pm$0.07} & $$8.5 & --  \\ 
J182854.87+011753.0 & -- & B & \multicolumn{2}{c}{0.21$\pm$0.04} & \multicolumn{2}{c}{0.24$\pm$0.07} & $$7.2 & --  \\ 
J182903.06+012331.0 & -- & B & 0.49$\pm$0.05 & 0.76$\pm$0.07 & \multicolumn{2}{c}{--} & $$7.5 & --  \\ 
J182905.07+012309.0 & -- & B & 0.26$\pm$0.04 & 0.31$\pm$0.05 & \multicolumn{2}{c}{--} & $>$6.5 & --  \\ 
J182910.17+012559.5 & SSTc2d$~$J182910.2+012560 & B & 2.70$\pm$0.05 & 3.45$\pm$0.05 & \multicolumn{2}{c}{--} & $$9.3 & --  \\ 
J182911.94+012119.4 & -- & B & 0.36$\pm$0.03 & 0.51$\pm$0.06 & \multicolumn{2}{c}{--} & $$7.7 & --  \\ 
J182916.11+010437.5 & SSTSL2$~$J182916.10+010438.6 & B & \multicolumn{2}{c}{0.33$\pm$0.06} & 0.24$\pm$0.05 & 0.26$\pm$0.08 & $$8.0 & --  \\ 
J182918.23+011757.7 & SSTc2d$~$J182918.2+011758 & B & 0.19$\pm$0.05 & 0.25$\pm$0.05 & \multicolumn{2}{c}{--} & $$7.3 & --  \\ 
J182926.71+012342.1 & SSTSL2$~$J182926.72+012342.4 & B & 0.17$\pm$0.05 & 0.25$\pm$0.06 & \multicolumn{2}{c}{--} & $$6.6 & --  \\ 
J182930.71+010048.3 & PMN$~$1829+0101 & B & 3.87$\pm$0.05 & 7.44$\pm$0.10 & \multicolumn{2}{c}{--} & $$8.7 & --  \\ 
J182933.07+011716.3 & GFM$~$11 & YSO & \multicolumn{2}{c}{0.19$\pm$0.04} & 0.27$\pm$0.05 & 0.33$\pm$0.06 & $>$6.6 & Class III  \\ 
J182935.02+011503.2 & DCE08-210$~$5 & B & 0.14$\pm$0.05 & 0.20$\pm$0.04 & \multicolumn{2}{c}{--} & $>$6.3 & --  \\ 
J182936.50+012317.0 & SSTc2d$~$J182936.5+012317 & B & 0.14$\pm$0.04 & 0.26$\pm$0.05 & \multicolumn{2}{c}{--} & $>$6.4 & --  \\ 
J182944.07+011921.1 & NVSS$~$182944+011920 & B & 1.41$\pm$0.04 & 1.74$\pm$0.04 & \multicolumn{2}{c}{--} & $>$7.2 & --  \\ 
J182948.83+010647.4 & SSTc2d$~$J182948.8+010648 & B & \multicolumn{2}{c}{0.35$\pm$0.05} & \multicolumn{2}{c}{0.63$\pm$0.07} & $$7.5 & -- \\ 
J182949.50+011955.8 & -- & B & 1.96$\pm$0.07 & 2.40$\pm$0.07 & \multicolumn{2}{c}{--} & $$7.6 & --  \\ 
J182951.04+011533.8 & ETC$~$8 & B & 0.35$\pm$0.06 & 0.59$\pm$0.05 & \multicolumn{2}{c}{--} & $$8.0 & --  \\ 
J182957.89+011246.0 & EC$~$95A &  \multirow{3}{*}{YSO} & 0.26$\pm$0.05 & 1.18$\pm$0.04 & \multicolumn{2}{c}{--} & $$8.3 & \multirow{3}{*}{P-HAeBe} \\ 
J182957.89+011246.0 & EC$~$95B &   & 0.16$\pm$0.04 & 1.17$\pm$0.04 & \multicolumn{2}{c}{--} & $$8.4 &   \\ 
J182957.89+011246.0 & EC$~$95C$^2$ &   & --  & -- & 0.86$\pm$0.19 &  3.68$\pm$0.10 &   $>$7.4  &  \\ 
J183000.65+011340.0 & GFM$~$65A & \multirow{2}{*}{YSO} & 0.26$\pm$0.05 & 0.50$\pm$0.04 & \multicolumn{2}{c}{--} & $>$6.7 & \multirow{2}{*}{Class III}  \\ 
J183000.65+011340.0 & GFM$~$65B &  & 0.22$\pm$0.05 & 0.57$\pm$0.11 & \multicolumn{2}{c}{--} & $$6.4 &    \\ 
J183004.62+012234.1 & GFM$~$70 & B & 0.41$\pm$0.05 & 0.42$\pm$0.05 & \multicolumn{2}{c}{--} & $>$6.6 & -- \\ 
J182952.73-015159.9 & -- & B & 0.20$\pm$0.05 & 0.26$\pm$0.07 & \multicolumn{2}{c}{--} & $$6.6 & --  \\ 
\cutinhead{W40}
J183044.11-020145.6 & 2M18304408--0201458 & B & 1.65$\pm$0.06 & 2.15$\pm$0.06 & \multicolumn{2}{c}{--} & $$7.9 & --  \\ 
J183114.82-020350.1 & KGF$~$36 & Star & 0.41$\pm$0.08 & 0.48$\pm$0.05 & 0.36$\pm$0.09 & 0.48$\pm$0.08 & $$7.3 & -- \\ 
J183118.68-015455.9 & -- & B & 0.43$\pm$0.10 & 0.52$\pm$0.06 & \multicolumn{2}{c}{1.14$\pm$0.18} & $$7.0 & --  \\ 
J183122.32-020619.6 & KGF$~$82 & YSO & \multicolumn{2}{c}{0.41$\pm$0.05} & \multicolumn{2}{c}{0.26$\pm$0.06} & $$7.6 & Class III  \\ 
J183123.62-020535.8 & KGF$~$97 & YSO & 0.10$\pm$0.05 & 1.21$\pm$0.05 & \multicolumn{2}{c}{--} & $$7.9 & Class III  \\ 
J183126.02-020517.0 & KGF$~$122 & YSO & 0.20$\pm$0.05 & 0.91$\pm$0.06 & \multicolumn{2}{c}{--} & $$8.0 & Class II \\ 
J183127.45-020512.0 & KGF$~$133 & YSO & 0.45$\pm$0.07 & 0.51$\pm$0.06 & \multicolumn{2}{c}{2.40$\pm$0.11} & $$7.7 & Class II/III  \\ 
J183127.65-020509.7 & KGF$~$138 & YSO & \multicolumn{2}{c}{0.35$\pm$0.06} & \multicolumn{2}{c}{--} & $>$6.5 & HAeBe  \\ 
\enddata
\tablenotetext{1}{GBS-VLA stands for Gould's Belt Very Large Array Survey \citep{Ortiz_15}. }
\tablenotetext{2}{Data corresponding to EC~95c were taken as part of projects BL160 and BD155, and are shown here for completeness. }
\tablenotetext{3}{Because most of the sources show significant flux variations, this value correspond to the maximum  brightness temperature }
\end{deluxetable}
\clearpage


\clearpage
\floattable
\begin{deluxetable}{ccccccccc}
\tablecolumns{9}
\tablewidth{0pt}
\tablecolumns{5}
\tablecaption{Measured positions of EC~95 \label{tab:ec95}}
\tablehead{ Julian Day  &Project$^{1}$ & $\alpha$ (J2000.0) & $\sigma_\alpha$ & $\delta$ (J2000.0) & $\sigma_\delta$    \\
}
\startdata
\sidehead{EC 95A}
2454800.39885 & BL160 & 18  29  57.89186638 & 0.00000180 & 1  12  46.110101 & 0.000069 \\ 
2454890.14136 & BL160 & 18  29  57.89217322 & 0.00000048 & 1  12  46.106940 & 0.000018 \\ 
2455171.38315 & BL160 & 18  29  57.89222331 & 0.00000098 & 1  12  46.095333 & 0.000041 \\ 
2455268.11855 & BL160 & 18  29  57.89247995 & 0.00000202 & 1  12  46.092081 & 0.000067 \\ 
2455356.87555 & BL160 & 18  29  57.89242962 & 0.00000117 & 1  12  46.089683 & 0.000054 \\ 
2455936.29042 & BD155 & 18  29  57.89251865 & 0.00000457 & 1  12  46.068868 & 0.000150 \\ 
2455937.28769 & BD155 & 18  29  57.89253227 & 0.00000298 & 1  12  46.068531 & 0.000082 \\ 
2456522.68545 & BD155 & 18  29  57.89239849 & 0.00000095 & 1  12  46.053528 & 0.000034 \\ 
2456524.67999 & BD155 & 18  29  57.89238944 & 0.00000228 & 1  12  46.053868 & 0.000114 \\ 
2456538.70634 & BL175 & 18  29  57.89232999 & 0.00000204 & 1  12  46.054528 & 0.000066 \\ 
2456720.20849 & BL175 & 18  29  57.89263275 & 0.00000445 & 1  12  46.049701 & 0.000122 \\ 
2456943.59865 & BL175 & 18  29  57.89233271 & 0.00000528 & 1  12  46.043694 & 0.000186 \\ 
2457507.02912 & BL175 & 18  29  57.89266020 & 0.00000104 & 1  12  46.032795 & 0.000036 \\
\sidehead{EC 95B}
2454457.31822 & BL156 & 18  29  57.89095609 & 0.00000120 & 1  12  46.107905 & 0.000038 \\ 
2454646.81935 & BL160 & 18  29  57.89095848 & 0.00000481 & 1  12  46.107242 & 0.000186 \\ 
2454724.60637 & BL160 & 18  29  57.89080948 & 0.00000083 & 1  12  46.105900 & 0.000029 \\ 
2454800.39885 & BL160 & 18  29  57.89088405 & 0.00000217 & 1  12  46.104416 & 0.000089 \\ 
2454890.14136 & BL160 & 18  29  57.89112095 & 0.00000388 & 1  12  46.103859 & 0.000138 \\ 
2454985.89100 & BL160 & 18  29  57.89106970 & 0.00000414 & 1  12  46.104177 & 0.000240 \\ 
2455074.64800 & BL160 & 18  29  57.89091190 & 0.00000082 & 1  12  46.103134 & 0.000032 \\ 
2455268.11855 & BL160 & 18  29  57.89131814 & 0.00000402 & 1  12  46.100962 & 0.000162 \\ 
2455356.87555 & BL160 & 18  29  57.89128563 & 0.00000363 & 1  12  46.101013 & 0.000176 \\ 
2455442.64072 & BL160 & 18  29  57.89116731 & 0.00000472 & 1  12  46.099877 & 0.000202 \\ 
2455936.29042 & BD155 & 18  29  57.89185545 & 0.00000614 & 1  12  46.091786 & 0.000156 \\ 
2456522.68545 & BD155 & 18  29  57.89246953 & 0.00000063 & 1  12  46.081657 & 0.000023 \\ 
2456524.67999 & BD155 & 18  29  57.89246863 & 0.00000421 & 1  12  46.081514 & 0.000137 \\ 
2456538.70634 & BL175 & 18  29  57.89244323 & 0.00000760 & 1  12  46.081859 & 0.000200 \\ 
2456720.20849 & BL175 & 18  29  57.89295395 & 0.00000877 & 1  12  46.078949 & 0.000243 \\ 
2456943.59865 & BL175 & 18  29  57.89292358 & 0.00000378 & 1  12  46.072521 & 0.000108 \\ 
2457084.21205 & BL175 & 18  29  57.89339762 & 0.00000179 & 1  12  46.068654 & 0.000071 \\ 
2457302.61352 & BL175 & 18  29  57.89339468 & 0.00000865 & 1  12  46.063242 & 0.000279 \\ 
2457391.30720 & BD155 & 18  29  57.89364862 & 0.00000016 & 1  12  46.060099 & 0.000006 \\ 
2457507.02912 & BL175 & 18  29  57.89389465 & 0.00000108 & 1  12  46.058656 & 0.000036 \\ 
\sidehead{EC 95C}
2454724.60637 & BL160 & 18  29  57.89856745 & 0.00000305 & 1  12  46.205651 & 0.000108 \\ 
2455936.29042 & BD155 & 18  29  57.89945356 & 0.00000060 & 1  12  46.166823 & 0.000019 \\
%
\enddata
\tablenotetext{1}{VLBA project code.}
\end{deluxetable}
\clearpage

\clearpage
\begin{deluxetable}{cccccccc}
\tabletypesize{\scriptsize}
\tablewidth{0pt}
\tablecolumns{6}
\tablecaption{Parallaxes and proper motions \label{tab:parallaxes}}
\tablehead{ GBS-VLA& Other identifier$^1$ & Parallax & $\mu_\alpha\cos\delta$ & $\mu_\delta$        &Distance \\
                   name      &                         & (mas)     & (mas yr$^{-1}$)               & (mas yr$^{-1}$)            & (pc)          \\
                   (1)           &           (2)         &    (3)       &   (4)                                 &   (5)                              &   (6)          \\
}
\startdata
J182933.07+011716.3  & GFM 11  & 2.313  $\pm$ 0.078  & 3.634  $\pm$ 0.050  & -8.864 $\pm$ 0.127  & 432.3  $\pm$ 14.6 \\
J182957.89+011246.0  &  EC 95    & 2.291  $\pm$ 0.038  & 3.599  $\pm$ 0.026  & -8.336  $\pm$ 0.030  & 436.4   $\pm$ 7.1 \\
J183000.65+011340.0  & GFM 65$^2$  & 2.638  $\pm$ 0.118  & 1.573  $\pm$ 0.070 & -6.513  $\pm$ 0.152  & 379.1  $\pm$ 17.0 \\
J183114.82-020350.1 &   KGF 36  & 2.302  $\pm$ 0.063 &   0.186  $\pm$ 0.053  & -6.726  $\pm$ 0.121  & 434.5   $\pm$ 11.8  \\
J183123.62-020535.8 &  KGF 97     & 2.186  $\pm$ 0.076  & -0.258   $\pm$ 0.058  & -7.514    $\pm$ 0.135  & 457.5  $\pm$ 16.0 \\
J183126.02-020517.0 &  KGF 122 & 2.372  $\pm$ 0.120   &  4.586   $\pm$ 0.074  & -7.946   $\pm$ 0.167  & 421.5  $\pm$ 21.4 \\
J183127.45-020512.0 &  KGF 133 & 2.385  $\pm$ 0.098  & -0.330   $\pm$ 0.049  & -7.746   $\pm$ 0.111  & 419.3  $\pm$ 17.3  \\
%
\enddata
\tablenotetext{1}{GFM = \cite{Giardino_2007};  EC= \cite{Eiroa_1995}, KGF = \cite{Kuhn_2010}. }
\tablenotetext{2}{Parallax solution could be affected by unmodelled binary motion. }
\end{deluxetable}
\clearpage
\clearpage
\floattable
\begin{deluxetable}{cccccccccccc}
\tabletypesize{\scriptsize}
\tablewidth{0pt}
\tablecolumns{6}
\tablecaption{Orbital elements of EC~95 \label{tab:ec95orbit_par}}
\tablehead{  Model  & a     & P   & $T_0$& $e$   & $\Omega$& $i$    & $\omega$       & $M_1$       & $M_2$      & $M_T$       \\
                              & (mas) & (yr)&      &     & ($^o$)  & ($^o$) & ($^o$)         & (M$_\odot$) & (M$_\odot$)& (M$_\odot$) \\
                  (1)         & (2)   & (3) & (4)  & (5) & (6)     & (7)    & (8)            & (9)         & (10)       & (11)        \\
}
\startdata
Full           & 28.9 $\pm$ 0.4  & 21.5 $\pm$ 1.5 & 2008.85 $\pm$ 2.0 & 0.397 $\pm$  0.001 & 124.8 $\pm$ 2.1  & 31.6 $\pm$ 0.9 & 477.5 $\pm$ 1.8 &  2.0 $\pm$ 0.2 & 2.3 $\pm$ 0.1 & 4.3 $\pm$ 0.2 \\ 
Rel. astr.     & 30.7 $\pm$ 1.4  & 23.1 $\pm$ 1.4 & 2009.08 $\pm$ 0.14 & 0.393 $\pm$ 0.011 & 136.2 $\pm$ 2.5 & 34.8 $\pm$ 2.0 & 475.3 $\pm$ 2.8 & --            & --            & 4.5 $\pm$ 0.2 \\ 
\enddata
\end{deluxetable}
\clearpage


\clearpage
\begin{deluxetable*}{lcccc}
\tabletypesize{\scriptsize}
\tablewidth{0pt}
\tablecolumns{5}
\tablecaption{Measured source positions \label{tab:positions}}
\tablehead{ Julian Day & $\alpha$ (J2000.0) & $\sigma_\alpha$ & $\delta$ (J2000.0) & $\sigma_\delta$    \\
}
\startdata
\cutinhead{SSTc2d$~$J182910.2+012560 } %
2456537.71083 & 18  29  10.18111439 & 0.00000048 & 1  25  59.593986 & 0.000016 \\ 
2456718.21396 & 18  29  10.18111515 & 0.00000048 & 1  25  59.593964 & 0.000016 \\ 
2456938.61153 & 18  29  10.18111964 & 0.00000046 & 1  25  59.594012 & 0.000016 \\ 
2457081.21921 & 18  29  10.18110324 & 0.00000063 & 1  25  59.594051 & 0.000020 \\ 
2457300.62039 & 18  29  10.18110850 & 0.00000051 & 1  25  59.594058 & 0.000017 \\ 
\cutinhead{SSTc2d$~$J182918.2+011758 } %
2456537.71083 & 18  29  18.23057894 & 0.00000623 & 1  17  57.783157 & 0.000237 \\ 
2456718.21396 & 18  29  18.23057676 & 0.00000432 & 1  17  57.782977 & 0.000165 \\ 
2456938.61153 & 18  29  18.23060031 & 0.00000827 & 1  17  57.783177 & 0.000266 \\ 
2457081.21921 & 18  29  18.23056481 & 0.00000605 & 1  17  57.782967 & 0.000199 \\ 
2457300.62039 & 18  29  18.23058354 & 0.00000919 & 1  17  57.783323 & 0.000305 \\ 
\cutinhead{GBS-VLA~J182903.06+012331.0 } %
2456537.71083 & 18  29  03.06984956 & 0.00000319 & 1  23  31.087749 & 0.000103 \\ 
2456718.21396 & 18  29  03.06986038 & 0.00000339 & 1  23  31.087530 & 0.000110 \\ 
2456938.61153 & 18  29  03.06986699 & 0.00000383 & 1  23  31.087498 & 0.000101 \\ 
2457081.21921 & 18  29  03.06987411 & 0.00000425 & 1  23  31.087780 & 0.000116 \\ 
2457300.62039 & 18  29  03.06987273 & 0.00000382 & 1  23  31.087686 & 0.000132 \\ 
\cutinhead{SSTc2d$~$J182936.5+012317 } %
2456537.71083 & 18  29  36.50110691 & 0.00000859 & 1  23  17.076353 & 0.000169 \\ 
2456718.21396 & 18  29  36.50114312 & 0.00000627 & 1  23  17.075662 & 0.000226 \\ 
2456938.61153 & 18  29  36.50113297 & 0.00000496 & 1  23  17.076045 & 0.000161 \\ 
2457081.21921 & 18  29  36.50112373 & 0.00001097 & 1  23  17.075484 & 0.000410 \\ 
2457300.62039 & 18  29  36.50110998 & 0.00000771 & 1  23  17.076210 & 0.000196 \\ 
\cutinhead{GBS-VLA~J182905.07+012309.0 } %
2456537.71083 & 18  29  05.08095263 & 0.00000445 & 1  23  09.149373 & 0.000137 \\ 
2456718.21396 & 18  29  05.08096153 & 0.00000384 & 1  23  09.149402 & 0.000155 \\ 
2457081.21921 & 18  29  05.08094750 & 0.00000559 & 1  23  09.149718 & 0.000160 \\ 
2457300.62039 & 18  29  05.08094627 & 0.00000762 & 1  23  09.150555 & 0.000146 \\ 
\cutinhead{SSTSL2$~$J182926.72+012342.4 } %
2456537.71083 & 18  29  26.71049982 & 0.00000918 & 1  23  42.131201 & 0.000268 \\ 
2456718.21396 & 18  29  26.71050367 & 0.00001050 & 1  23  42.130807 & 0.000385 \\ 
\cutinhead{GBS-VLA~J182911.94+012119.4 } %
2456537.71083 & 18  29  11.94832584 & 0.00000362 & 1  21  19.484862 & 0.000107 \\ 
2456718.21396 & 18  29  11.94830530 & 0.00000283 & 1  21  19.485004 & 0.000112 \\ 
2456938.61153 & 18  29  11.94832429 & 0.00000313 & 1  21  19.484997 & 0.000106 \\ 
2457081.21921 & 18  29  11.94830236 & 0.00000478 & 1  21  19.485052 & 0.000132 \\ 
2457300.62039 & 18  29  11.94833353 & 0.00000291 & 1  21  19.484955 & 0.000089 \\ 
\cutinhead{NVSS$~$182944+011920 } %
2456537.71083 & 18  29  44.07658313 & 0.00000075 & 1  19  21.164119 & 0.000025 \\ 
2456718.21396 & 18  29  44.07657402 & 0.00000085 & 1  19  21.164280 & 0.000028 \\ 
2456938.61153 & 18  29  44.07659031 & 0.00000068 & 1  19  21.164227 & 0.000023 \\ 
2457081.21921 & 18  29  44.07657708 & 0.00000112 & 1  19  21.164286 & 0.000035 \\ 
2457300.62039 & 18  29  44.07655805 & 0.00000085 & 1  19  21.164512 & 0.000029 \\ 
\cutinhead{PMN$~$1829+0101 } %
2456538.70634 & 18  29  30.72388371 & 0.00000033 & 1  0  48.005138 & 0.000010 \\ 
2456720.20849 & 18  29  30.72391467 & 0.00000085 & 1  0  48.005243 & 0.000022 \\ 
2456943.59865 & 18  29  30.72388420 & 0.00000045 & 1  0  48.004833 & 0.000016 \\ 
\cutinhead{DCE08-210$~$5 } %
2456538.70634 & 18  29  35.02394353 & 0.00001739 & 1  15  03.254608 & 0.000371 \\ 
2456720.20849 & 18  29  35.02397943 & 0.00001364 & 1  15  03.252858 & 0.000403 \\ 
2457507.02912 & 18  29  35.02401622 & 0.00000701 & 1  15  03.253580 & 0.000198 \\ 
\cutinhead{GFM$~$65 } %
2456720.20849 & 18  30  00.65283020 & 0.00000429 & 1  13  40.065067 & 0.000136 \\ 
2456943.59865 & 18  30  00.65256389 & 0.00000712 & 1  13  40.061403 & 0.000205 \\ 
2457302.61352 & 18  30  00.65266340 & 0.00000277 & 1  13  40.054815 & 0.000088 \\ 
second source: \\
2456720.20849 & 18  30  00.65296360 & 0.00002331 & 1  13  40.066599 & 0.000638 \\ 
2456943.59865 & 18  30  00.65289182 & 0.00000696 & 1  13  40.061476 & 0.000276 \\ 
\cutinhead{ETC$~$8 } %
2456720.20849 & 18  29  51.04143717 & 0.00000565 & 1  15  33.871396 & 0.000160 \\ 
2456943.59865 & 18  29  51.04141126 & 0.00000725 & 1  15  33.870733 & 0.000178 \\ 
2457084.21205 & 18  29  51.04142997 & 0.00000353 & 1  15  33.870437 & 0.000131 \\ 
2457302.61352 & 18  29  51.04141829 & 0.00000270 & 1  15  33.870371 & 0.000080 \\ 
2457507.02912 & 18  29  51.04142929 & 0.00000406 & 1  15  33.870259 & 0.000106 \\ 
\cutinhead{KGF$~$122 } %
2456539.70380 & 18  31  26.01996029 & 0.00000276 & -2  5  17.086151 & 0.000086 \\ 
2457102.16179 & 18  31  26.02073064 & 0.00000306 & -2  5  17.098461 & 0.000137 \\ 
2457285.66188 & 18  31  26.02057197 & 0.00001863 & -2  5  17.102963 & 0.000368 \\ 
2457448.21552 & 18  31  26.02087306 & 0.00001310 & -2  5  17.105731 & 0.000405 \\ 
\cutinhead{2M18304408--0201458 } %
2456539.70380 & 18  30  44.11485642 & 0.00000199 & -2  1  45.688322 & 0.000051 \\ 
2456721.20565 & 18  30  44.11487547 & 0.00000271 & -2  1  45.689938 & 0.000071 \\ 
2456946.58954 & 18  30  44.11484733 & 0.00000336 & -2  1  45.688604 & 0.000087 \\ 
2457102.16179 & 18  30  44.11486486 & 0.00000140 & -2  1  45.688221 & 0.000038 \\ 
2457285.66188 & 18  30  44.11485685 & 0.00000236 & -2  1  45.688281 & 0.000057 \\ 
\cutinhead{KGF$~$138 } %
2457102.16179 & 18  31  27.65620135 & 0.00000709 & -2  5  09.799495 & 0.000178 \\ 
\cutinhead{KGF$~$97 } %
2456721.20565 & 18  31  23.62227407 & 0.00000250 & -2  5  35.868544 & 0.000074 \\ 
2456946.58954 & 18  31  23.62203557 & 0.00002812 & -2  5  35.873472 & 0.000761 \\ 
2457102.16179 & 18  31  23.62227217 & 0.00001073 & -2  5  35.876576 & 0.000314 \\ 
2457285.66188 & 18  31  23.62197113 & 0.00000699 & -2  5  35.880042 & 0.000161 \\ 
2457448.21552 & 18  31  23.62223545 & 0.00000309 & -2  5  35.883516 & 0.000086 \\ 
\cutinhead{KGF$~$36 } %
2456543.60141 & 18  31  14.82263201 & 0.00000386 & -2  3  50.149196 & 0.000131 \\ 
2456726.05521 & 18  31  14.82293762 & 0.00000369 & -2  3  50.152750 & 0.000106 \\ 
2457306.96324 & 18  31  14.82267359 & 0.00000980 & -2  3  50.163179 & 0.000371 \\ 
2457448.21552 & 18  31  14.82293818 & 0.00000571 & -2  3  50.166392 & 0.000155 \\ 
\cutinhead{GBS-VLA~J182952.73--015159.9 } %
2456541.69591 & 18  29  52.73464125 & 0.00001627 & -1  51  59.925315 & 0.000561 \\ 
2456723.19777 & 18  29  52.73465607 & 0.00002462 & -1  51  59.926572 & 0.000665 \\ 
2457309.09678 & 18  29  52.73469573 & 0.00001149 & -1  51  59.925989 & 0.000363 \\ 
\cutinhead{KGF$~$82 } %
2456543.60141 & 18  31  22.32975638 & 0.00000485 & -2  6  19.633463 & 0.000140 \\ 
2457306.96324 & 18  31  22.32894475 & 0.00000404 & -2  6  19.660373 & 0.000166 \\ 
\cutinhead{KGF$~$133 } %
2456543.60141 & 18  31  27.45984260 & 0.00000137 & -2  5  12.036684 & 0.000039 \\ 
2457306.96324 & 18  31  27.45978832 & 0.00000657 & -2  5  12.052156 & 0.000273 \\ 
2457448.21552 & 18  31  27.46007650 & 0.00000568 & -2  5  12.056771 & 0.000139 \\ 
\cutinhead{GBS-VLA~J183118.68--015455.9 } %
2456706.16961 & 18  31  18.68250486 & 0.00001724 & -1  54  56.073788 & 0.000262 \\ 
2457084.21205 & 18  31  18.68250948 & 0.00001509 & -1  54  56.073322 & 0.000371 \\ 
2457302.61352 & 18  31  18.68249661 & 0.00000858 & -1  54  56.072971 & 0.000207 \\ 
\cutinhead{GBS-VLA~J182949.50+011955.8 } %
2456541.69877 & 18  29  49.50633251 & 0.00000273 & 1  19  55.885107 & 0.000090 \\ 
2456723.20062 & 18  29  49.50632851 & 0.00000189 & 1  19  55.885721 & 0.000123 \\ 
2457309.09678 & 18  29  49.50633154 & 0.00000120 & 1  19  55.885384 & 0.000068 \\ 
\cutinhead{GFM$~$70 } %
2456723.20062 & 18  30  04.62941667 & 0.00000502 & 1  22  34.131415 & 0.000132 \\ 
2457309.09678 & 18  30  04.62940928 & 0.00000397 & 1  22  34.131168 & 0.000121 \\ 
\cutinhead{GFM$~$11 } %
2456543.69416 & 18  29  33.07249309 & 0.00000229 & 1  17  16.360959 & 0.000098 \\ 
2456726.19504 & 18  29  33.07290417 & 0.00000527 & 1  17  16.356120 & 0.000275 \\ 
2457507.02912 & 18  29  33.07340833 & 0.00000593 & 1  17  16.337928 & 0.000238 \\ 
\cutinhead{SSTSL2$~$J182916.10+010438.6 } %
2456543.69416 & 18  29  16.11947301 & 0.00000517 & 1  4  37.589379 & 0.000201 \\ 
2456726.19504 & 18  29  16.11946492 & 0.00000667 & 1  4  37.589438 & 0.000220 \\ 
2457307.60223 & 18  29  16.11947017 & 0.00000623 & 1  4  37.589495 & 0.000238 \\ 
\cutinhead{GBS-VLA~J182854.46+011823.7 } %
2456749.07794 & 18  28  54.46499411 & 0.00000049 & 1  18  23.813820 & 0.000015 \\ 
2457281.61856 & 18  28  54.46500889 & 0.00000023 & 1  18  23.813619 & 0.000008 \\ 
\cutinhead{GBS-VLA~J182854.44+011859.7 } %
2457281.61856 & 18  28  54.44344643 & 0.00000226 & 1  18  59.737811 & 0.000112 \\ 
\cutinhead{GBS-VLA~J182854.87+011753.0 } %
2456749.07794 & 18  28  54.87254346 & 0.00000603 & 1  17  53.051049 & 0.000192 \\ 
2457281.61856 & 18  28  54.87253760 & 0.00000534 & 1  17  53.051239 & 0.000208 \\ 
\cutinhead{SSTc2d$~$J182948.8+010648 } %
2456749.09635 & 18  29  48.82981795 & 0.00000628 & 1  6  47.450268 & 0.000181 \\ 
2457281.61856 & 18  29  48.82980672 & 0.00000240 & 1  6  47.450032 & 0.000084 \\ 
\enddata
\end{deluxetable*}
\clearpage

\clearpage

\end{document}